\documentclass[reprint, amsmath,amssymb,aps,floatfix]{revtex4-2}
\usepackage{hyperref}
\usepackage{graphicx}
\usepackage[usenames,dvipsnames,svgnames,table]{xcolor}
\usepackage{soul} 
\usepackage{gensymb}

\begin{document}

\title{Enhanced nonlinear optomechanics in a coupled-mode photonic crystal device}
\date{\today}
\author{Roel Burgwal}
\affiliation{Department of Applied Physics and Eindhoven Hendrik Casimir Institute, Eindhoven University of Technology, P.O. Box 513, 5600 MB Eindhoven, The Netherlands}
\affiliation{Center for Nanophotonics, AMOLF, Science Park 104, 1098 XG Amsterdam, The Netherlands}
\author{Ewold Verhagen}
\affiliation{Department of Applied Physics and Eindhoven Hendrik Casimir Institute, Eindhoven University of Technology, P.O. Box 513, 5600 MB Eindhoven, The Netherlands}
\affiliation{Center for Nanophotonics, AMOLF, Science Park 104, 1098 XG Amsterdam, The Netherlands}

\begin{abstract}
    The nonlinear component of the optomechanical interaction between light and mechanical vibration promises many exciting classical and quantum mechanical applications, but is generally weak. Here we demonstrate enhancement of nonlinear optomechanical measurement of mechanical motion by using pairs of coupled optical and mechanical modes in a photonic crystal device. In the same device we show linear optomechanical measurement with a strongly reduced input power and reveal how both enhancements are related. Our design exploits anisotropic mechanical elasticity to create strong coupling between mechanical modes while not changing optical properties. Additional thermo-optic tuning of the optical modes is performed with an auxiliary laser and a thermally-optimised device design. We envision broad use of this enhancement scheme in multimode phonon lasing, two-phonon heralding and eventually nonlinear quantum optomechanics.
\end{abstract}
\maketitle

\section*{Introduction}
The field of cavity optomechanics studies the interaction between a light field and mechanical vibration. On the one hand, the optomechanical interaction imprints the mechanical motion onto the light field, enabling extremely precise optical detection of position and spurring the development of highly precise sensors that approach and even evade fundamental measurement limits set by quantum mechanics \cite{Mason2019,Muhonen2019,magrini2021}. At the same time, the light field can be used to manipulate the state of the mechanical resonator, which has allowed the creation of mechanical quantum states for use in quantum information technology, as information storage or as tool in the conversion of superconducting microwave qubits to optical qubits \cite{wallucks2020,bochmann2013,andrews2014,mirhosseini2020,jiang2020}.

Many especially exciting applications have been envisioned that exploit nonlinear interaction between the light field and mechanical modes. The optomechanical interaction is in fact inherently nonlinear, but for current systems the linear component is dominating for quantum-level mechanical motion. A sufficiently strong nonlinearity would open up possibilities such as measurement-based non-classical state generation \cite{Brawley2016}, energy-squeezed states \cite{ma2021}, quantum non-demolition (QND) measurement of phonon number \cite{Thompson2008,Clerk2010} or the photon-blockade effect \cite{Rabl2011,Stannigel2012}. Such effects become apparent in the single-photon strong coupling (SPSC) limit $g_0/\kappa > 1$, where $g_0$ is the optomechanical vacuum coupling rate, and $\kappa$ is the decay rate of the optical resonator, but this limit is still out of reach of current state-of-the-art devices.

Nonlinear optomechanical effects can be enhanced in a system of two coupled optical modes, both optomechanically coupled to one mechanical mode, often referred to as the membrane-in-the-middle (MIM) system \cite{Thompson2008}. This enhancement is particularly interesting for systems that approach the SPSC regime, as it makes nonlinear quantum effects more pronounced \cite{ludwig2012,Stannigel2012}. However, enhancement of the nonlinearity in such a multimode system over its magnitude in a comparable single-mode system is only possible when fulfilling two requirements on the system parameters, namely that the coupling rate between the two optical modes $J_\mathrm{O}$ has to equal half the mechanical frequency $\Omega$ ($J_\mathrm{O} = \Omega/2$), and that the mechanical frequency is larger than the optical decay rate $\kappa$ ($\Omega > \kappa$), i.e. the sideband resolution condition \cite{burgwal2020}.

There have been many realisations of the MIM and related systems, in membranes \cite{Thompson2008,Lee2015}, microtoroids \cite{Grudinin2010,Hill2013}, photonic crystals \cite{healey2015,Paraiso2015,Fang2017}, ultracold atoms \cite{Purdy2010} and levitated particles \cite{bullier2021}. However, the systems in which nonlinear transduction was studied did not have the required inter-mode optical coupling ($J_\mathrm{O} \approx \Omega/2$) and optical decay rate ($\Omega > \kappa$) to exhibit nonlinear effects that were enhanced above the intrinsic optomechanical nonlinearity and a direct experimental comparison is lacking. Realising the required parameters in a photonic crystal system would be specifically interesting, as these are receiving strong attention due to their large optomechanical coupling, small footprint, and compatibility with cryogenic operation \cite{riedinger2018,mirhosseini2020,jiang2020}. Importantly, such multimode photonic crystal systems would also be useful to enhance linear transduction per input optical power \cite{Dobrindt2010}. 

\begin{figure*}
    \centering
    \includegraphics[width=\textwidth]{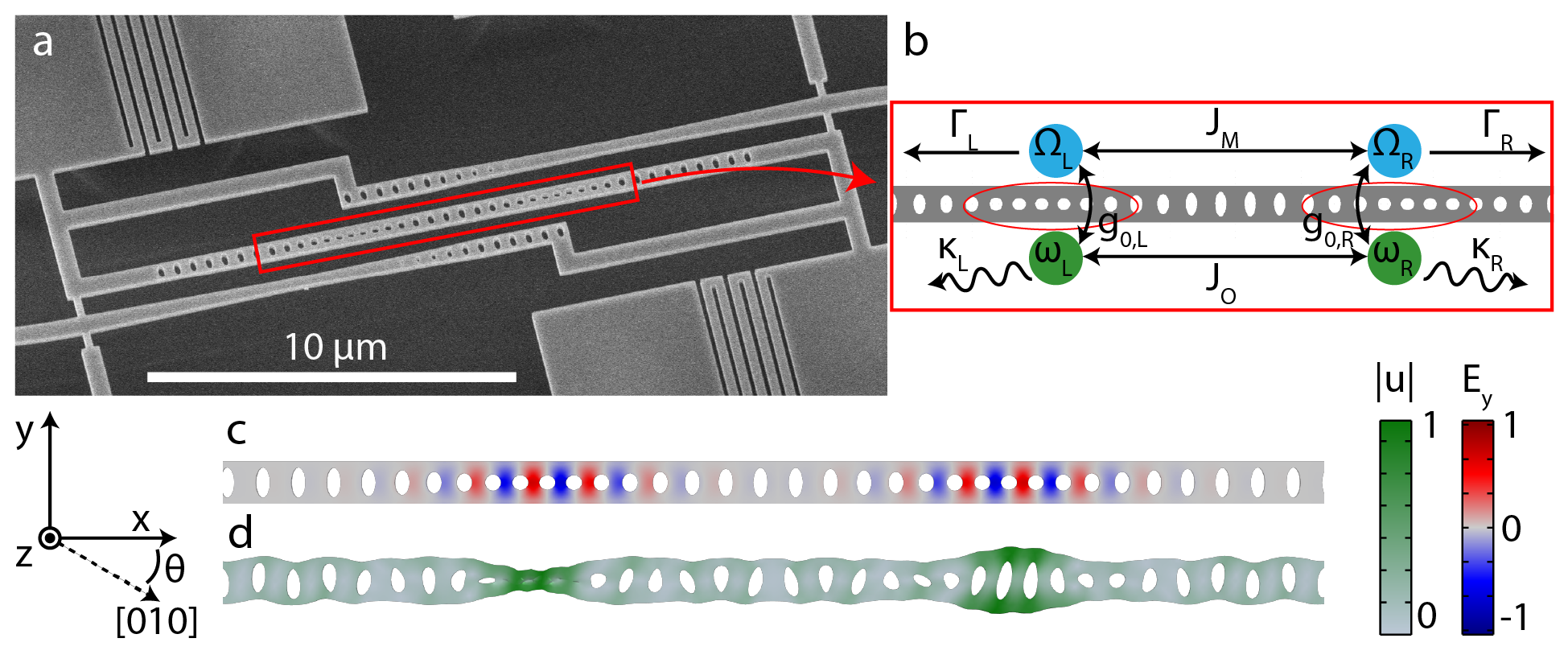}
    \caption{\textbf{Coupled-mode device design} a) A scanning electron microscope image of a fabricated device with nanobeam, waveguides and support structure. b) Schematic of the mode frequencies and coupling rates in the double-cavity nanobeam. c) The simulated $y$-component of the electrical field for the even optical supermode. d) The displacement ($|u|$) profile of a simulated mechanical supermode, which has odd symmetry. Note that the depicted amplitude of motion is largely exaggerated. An angle of $\theta =20\degree$ to the crystal axis was used for this simulation. }
    \label{fig1}
\end{figure*}

Here, we describe a coupled-mode optomechanical crystal device fulfilling all the above requirements, and use it to demonstrate for the first time enhanced nonlinear optomechanical coupling in a direct comparison to a single-mode configuration in the same device. To do so, we measure nonlinear transduction of thermomechanical motion in a coupled-mode device, where one of the optical modes can be selectively and actively detuned to switch between a single and coupled-mode configuration. Additionally, we quantify the enhancement of linear transduction with respect to input power that is also present in these systems, thus demonstrating two main advantages of the coupled-mode system. Our device shows strong coupling of optical and mechanical modes, for which we explored the use of mechanical anisotropy of silicon to tune mechanical properties without affecting the optical properties of the device. Finally, post-fabrication tuning of the optical modes, needed to correct inevitable fabrication imperfections, is achieved using thermal tuning with a laser as heat source and a thermally-optimised device design.

\section*{Results}

\subsection*{Model} \label{sec:model}
In the MIM system, two optical modes with annihilation operators $\hat{a}_\mathrm{L},\hat{a}_\mathrm{R}$ and frequency $\omega$ couple to each other with rate $J_\mathrm{O}$, and optomechanically to a mechanical mode with position (momentum) operator $\hat{x}\;(\hat{p})$ with vacuum coupling rates $g_\mathrm{L},g_\mathrm{R}$. This creates a Hamiltonian (setting $\hbar=1$) \cite{Kalaee2016}:
\begin{multline} \label{eq:ham}
    \hat{H} = (\omega + g_\mathrm{L}\hat{x})\hat{a}^\dagger_\mathrm{L} \hat{a}_\mathrm{L} + (\omega + g_\mathrm{R}\hat{x})\hat{a}^\dagger_\mathrm{R} \hat{a}_\mathrm{R} - J_\mathrm{O}(\hat{a}^\dagger_\mathrm{L}\hat{a}_\mathrm{R} + \hat{a}^\dagger_\mathrm{R}\hat{a}_\mathrm{L}) \\ + \hat{H}_\mathrm{m},
\end{multline}
where $\hat{H}_\mathrm{m} = \hat{p}^2/(2m) + (1/2)m\Omega_m^2\hat{x}^2$ is the mechanical Hamiltonian with $m$ the mass and $\Omega$ the mechanical frequency. By moving to a basis of odd and even optical supermodes $\hat{a}_\mathrm{e(o)} = 1/\sqrt{2}(\hat{a}_\mathrm{L} \pm \hat{a}_\mathrm{R}$), the Hamiltonian can be written as:
\begin{multline} \label{eq:superm_ham}
    \hat{H} = (\omega-J_\mathrm{O})\hat{a}^\dagger_\mathrm{e}\hat{a}_\mathrm{e} + (\omega+J_\mathrm{O})\hat{a}^\dagger_\mathrm{o}\hat{a}_\mathrm{o} \\+ \hat{x}\frac{g_\mathrm{L} + g_\mathrm{R}}{2}(\hat{a}^\dagger_\mathrm{e} \hat{a}_\mathrm{e} + \hat{a}^\dagger_\mathrm{o} \hat{a}_\mathrm{o}) \\ + \hat{x}\frac{g_\mathrm{L} - g_\mathrm{R}}{2}(\hat{a}^\dagger_\mathrm{e} \hat{a}_\mathrm{o} + \hat{a}^\dagger_\mathrm{o}\hat{a}_\mathrm{e}) + \hat{H}_\mathrm{m},
\end{multline}
describing a new system with two optical eigenmodes separated in frequency by $2J_\mathrm{O}$. We consider $g_\mathrm{L}=g_\mathrm{R}$ ($g_\mathrm{L}=-g_\mathrm{R}$), in which situation we call the mechanical mode described by $\hat{x}$ \textit{even} (\textit{odd}). For an even mode, we have optomechanical interaction terms of the form $\hat{x}\hat{a}^\dagger_\mathrm{e(o)}\hat{a}_\mathrm{e(o)}$, of similar form to the canonical, single-mode optomechanical system. However, for an odd mode, interaction terms are of the form $\hat{x}\hat{a}^\dagger_\mathrm{e(o)}\hat{a}_\mathrm{o(e)}$, so-called \textit{cross-mode} interactions. 

For an odd mechanical mode, under the condition of slow mechanical motion $\Omega \ll J_\mathrm{O}$, it is possible to diagonalise the Hamiltonian to isolate the quadratic coupling:
\begin{equation} \label{eq:qua_ham}
    \hat{H}_\mathrm{int} = \frac{(g_\mathrm{L} + g_\mathrm{R})^2}{8J}\hat{x}^2(\hat{a}^\dagger_\mathrm{o}\hat{a}_\mathrm{o} - \hat{a}^\dagger_\mathrm{e} \hat{a}_\mathrm{e}),
\end{equation}
which promises a large nonlinear interaction for small $J_\mathrm{O}$ \cite{Thompson2008}. However, it was found early on that this form fails to capture remaining linear interaction \cite{Miao2009,Yanay2016}, which precludes many applications such as a measurement of phonon number without reaching the SPSC limit. Moreover, it was shown that, in order for nonlinear interaction to be enhanced, sideband resolution $\Omega > \kappa$ and a specific optical coupling rate of $J_\mathrm{O}\approx \Omega/2$ is required \cite{Stannigel2012,ludwig2012,burgwal2020}.

To describe both linear and nonlinear transduction fully, we solve the Langevin equations of motion derived from the Hamiltonian in Eq. \ref{eq:ham}, 
with operators replaced by their expectation values, $a=\langle \hat{a}\rangle$. The equations are solved perturbatively to second order to capture nonlinear effects, working in a frame rotating with the optical input field frequency $\omega_\mathrm{in}$. The perturbative approach assumes that the mechanical motion is small, i.e. $\sqrt{\langle \hat{x}^2 \rangle } < \kappa/g_0$, which is true for thermal motion in most of current optomechanical devices. 

Using these equations, expressions can be derived (see Methods for details) for the photocurrent $I$ power spectral density (PSD) $S_\mathrm{II}[\omega]$ of heterodyne detection of light reflected from the optomechanical cavity, which can be compared to spectrum analyser measurements described below. The mechanical mode is assumed to be odd ($g_\mathrm{R}=-g_\mathrm{L}=g$) and driven only by the thermal environment, while only the left optical mode is probed. Then, for linear transduction in a single-mode device, the heterodyne PSD can be approximated by (see SI for conditions)
\begin{equation} \label{eq:lintransfulldet}
     S_\mathrm{II}^\mathrm{lin}[\Omega] = \kappa_\mathrm{ex,L}^2 n_\mathrm{th} g^2 n_\mathrm{in} \left|\frac{\chi(-\Omega)}{\Delta_\mathrm{L}(-\Omega+\Delta_\mathrm{L})}\right|^2,
\end{equation}
while for the coupled-mode system, it reads
\begin{multline} \label{eq:lintransfulltun}
    S_\mathrm{II}^\mathrm{lin}[\Omega] = \frac{ \kappa_\mathrm{ex,L}^2 n_\mathrm{th}g^2 n_\mathrm{in}}{4} 
    \\ \left| \frac{\chi(-\Omega)}{(-\Omega + \Delta_\mathrm{L} + J_\mathrm{O})(\Delta_\mathrm{L}-J_\mathrm{O})} \right|^2,
\end{multline}
where we have introduced the complex detuning $\Delta_\mathrm{L(R)} = (\omega_\mathrm{in}-\omega_\mathrm{L(R)}) + i\kappa_\mathrm{L(R)}/2$, which contains as a real part the left (right) laser-cavity detuning, and as imaginary part contains the optical decay rate $\kappa_\mathrm{L(R)}$, and $\kappa_\mathrm{ex,L(R)}$ is the outcoupling rate of the cavities to their respective read-out ports. The average amount of thermal phonons $n_\mathrm{th}$ can be expressed as $n_\mathrm{th} \approx k_\mathrm{B}T/(\hbar \Omega)$, with $k_\mathrm{B}$ the Boltzmann constant, $T$ the temperature, and $\chi(\omega) = 2\sqrt{\Gamma}\Omega/(\Omega^2-\omega^2-i\omega\Gamma)$, the mechanical susceptibility, with $\Gamma$ the mechanical decay rate, ignoring here optomechanical backaction effects on the mechanical mode at high powers for simplicity. Finally, $n_\mathrm{in}$ is the amount of photons per second in the optical input field. In the coupled system, for optimal $\mathrm{Re}(\Delta) = J_\mathrm{O}$ and $J_\mathrm{O}=\Omega/2$, both terms in the denominator can be minimised simultaneously and transduction at the mechanical frequency $\Omega$ reaches:
\begin{equation} 
    \max(S_\mathrm{II}^\mathrm{lin}) = \frac{16 \kappa_\mathrm{ex,L}^2 n_\mathrm{th} n_\mathrm{in} g^2}{\Gamma \kappa^4}.
\end{equation}
Compared to optimal linear transduction in a single-cavity system, where it is not possible to minimize both terms in the denominator simultaneously, that gives an enhancement of optomechanical sideband power of:
\begin{equation}
    \mathcal{E}^\mathrm{lin} = \frac{\max(S_\mathrm{II}^\mathrm{lin}[\Omega])_\mathrm{coupled}}{\max(S_\mathrm{II}^\mathrm{lin}[\Omega])_\mathrm{single}} =\left(\frac{\Omega}{\kappa}\right)^2.
\end{equation}
Thus, for equal optical input power $\mathcal{P}_\mathrm{in}=\hbar \omega_\mathrm{in} n_\mathrm{in}$, the coupled-mode system can improve linear optical read-out of mechanical motion. The creation of fluctuations in the cavity field through the optomechanical interaction can also be viewed as the inelastic scattering of light from the input frequency to sidebands at frequencies $\Omega$ lower or higher than $\omega_\mathrm{in}$. In this picture, the linear enhancement in coupled-mode systems can be regarded as using the two optical supermodes to achieve simultaneous resonance of both the input field and the optomechanically scattered sideband \cite{Dobrindt2010}. As a result, the intracavity photon number is larger in the coupled cavity case than in the single cavity pumped at equal power $\mathcal{P}_\mathrm{in}$.

Nonlinear optomechanical interaction manifests itself as fluctuations in the reflected light at twice the mechanical frequency. Such fluctuations can be calculated when solving the EOMs to second order, where they give a detector PSD that can be simplified for a single-mode system to
\begin{multline} \label{eq:nonlintransfulldet}
        S_\mathrm{II}^\mathrm{qua}[2\Omega] = \frac{\kappa_\mathrm{ex,L} n_\mathrm{th}^2 n_\mathrm{cav} g^4}{\pi} \\ \int_{-\infty}^\infty d\omega^\prime \left| \frac{\chi(\omega^\prime) \chi(-2\Omega-\omega^\prime)}{(-2\Omega + \Delta_\mathrm{L})(-2\Omega -\omega^\prime +\Delta_\mathrm{L})}  \right|^2,
\end{multline}
while for a coupled-mode system it reads
\begin{multline} \label{eq:nonlintransfulltun}
        S_\mathrm{II}^\mathrm{qua}[2\Omega] = \frac{\kappa_\mathrm{ex,L} n_\mathrm{th}^2 n_\mathrm{cav} g^4 }{2\pi} \\
    \int_{-\infty}^\infty d\omega^\prime \left| \frac{\chi(\omega^\prime)\chi(-2\Omega_\mathrm{m}-\omega^\prime)}{(-2\Omega_m + \Delta + J_\mathrm{O})(-2\Omega-\omega^\prime + \Delta -J_\mathrm{O})} \right|^2,
\end{multline}
where $n_\mathrm{cav}$ is the amount of photons in the cavity. For derivations and further details, see the SI. In contrast to the linear enhancement, the nonlinear enhancement persists when normalising to the amount of photons in the cavity, which is the limiting factor in many experiments \cite{riedinger2018,forsch2020,mirhosseini2020}. 
Again, the nonlinear transduction can be optimised in the coupled-mode system for the resonance condition $J_\mathrm{O}=\Omega/2$, $\mathrm{Re}(\Delta)=J_\mathrm{O}+\Omega$, where it reads: 
\begin{equation}
    \max(S_\mathrm{II}^\mathrm{qua}(2\Omega)) \approx \frac{64 \kappa_\mathrm{ex,L} n_\mathrm{th}^2 n_\mathrm{cav} g^4}{\Gamma \kappa^4}.
\end{equation}
Comparing this to optimal transduction in a single-cavity system, for which only one term in the denominator can be minimised, we find an enhancement of nonlinear transduction given by:
\begin{equation}
    \mathcal{E}^\mathrm{qua} = \frac{\max(S_\mathrm{II}^\mathrm{qua}[2\Omega])_\mathrm{coupled}}{\max(S_\mathrm{II}^\mathrm{qua}[2\Omega)])_\mathrm{single}} =2\left(\frac{\Omega}{\kappa}\right)^2.
\end{equation}
This factor captures the optimal enhancement of nonlinear transduction possible in the coupled-mode system, which we find limited by the degree of sideband resolution $\Omega/\kappa$ of the system. As we will discuss in more detail later, the minimisation of both terms in the denominator of Eq. \ref{eq:nonlintransfulltun} can be understood as simultaneous resonance of the linearly scattered (the first sideband) and nonlinearly scattered light (the second sideband) with one of the optical supermodes.

\subsection*{Coupled-mode design principle}
As a basis for our coupled-mode device, we use a one-dimensional optomechanical crystal nanobeam in which an optical and mechanical mode are co-localised in a defect or cavity region to create a large optomechanical coupling \cite{Chan2012}. Used often in recent quantum optomechanics experiments \cite{chan2011,bochmann2013,shomroni2019,wallucks2020,mirhosseini2020,jiang2020,riedinger2018,forsch2020}, this cavity is particularly attractive because of its large optomechanical coupling $g_0$, operation in sideband resolved regime $\Omega > \kappa$ and potential for ground-state initialisation in a cryogenic environment because of its high mechanical frequency ($\approx 5$ GHz). Building a coupled-mode system from such favorable single-cavity building blocks ensures best possible performance of the coupled system. Starting from this basis, we create two optomechanical cavities by writing two crystal defect regions in the same nanobeam (see Fig. \ref{fig1}a and b). Through overlap of the evanescent fields of the cavity modes, couplings between the two optical as well as the two mechanical modes are created, characterised by inter-cavity coupling rates $J_\mathrm{O}$ and $J_\mathrm{M}$, respectively. The mode frequencies (decay rates) are given by $\omega_i$ ($\kappa_i$) and $\Omega_i$ ($\Gamma_i$) for the optical and mechanical modes, respectively, where $i\in \{\mathrm{L},\mathrm{R}\}$ indicates the left and right cavities. Furthermore, we include next to our nanobeam two waveguides that allow us to couple to either the right or left cavity individually. These waveguides in turn connect to a dimpled, tapered optical fiber (see Fig. \ref{fig1}a) \cite{Groblacher2013}. The cavity-waveguide coupling rates are given by $\kappa_\mathrm{ex,i}$.

If the inter-cavity coupling exceeds the decay rates of the modes, as well as any possible frequency difference between the two modes, the local optical or mechanical modes hybridise into odd and even combinations of the left and right cavities, which are split in frequency by $2J_\mathrm{O}$ or $2J_\mathrm{M}$. Using finite-element method (FEM) simulations, we calculate the optical eigenmode frequencies of a nanobeam design and deduce $J_\mathrm{O}$ from the supermode frequency difference. In Fig. \ref{fig1}d, an example of a simulated optical supermode is plotted. As we require $2J_\mathrm{O}=\Omega$ for optimal enhancement of optomechanical effects, accurate control over the optical coupling rate is crucial. Coupling rates can be varied by changing the number and shape of the holes that make up the optomechanical crystal between the cavities, i.e. the coupling region.

After coupling region optimisation for optical coupling rate (see SI), the device design has a mechanical coupling rate that will typically not allow for strong coupling of the mechanical modes, as fabrication imperfections induce random frequency differences between the two mechanical modes that have to be overcome by a sufficiently large coupling rate. For independent tuning of the mechanical coupling rate, we exploit the anisotropy of the mechanical properties of the device material, monocrystalline silicon. By fabricating devices at an angle $\theta$ to the $\langle 010\rangle$ crystal axis, the mechanical properties can be varied while leaving optical properties unaltered.

We studied the behaviour of mechanical modes as a function of fabrication angle $\theta$ using FEM simulations (see SI for full details). For a non-zero angle, the $y$-symmetry of the system (orthogonal to beam axis, in plane) is broken due to anisotropy. Although the nanobeam cavities are designed to localise the $y$-symmetric (\textit{breathing}) mode, other modes exist at the same frequency with $y$-antisymmetry that are not confined by the defects. The introduction of nonzero $\theta$ mixes the breathing and $y$-antisymmetric modes and thus produces a new mode which can leak from the cavity more easily, resulting in a stronger effective $J_\mathrm{M}$ and the formation of supermodes, of which an example is plotted in Fig. \ref{fig1}d. Note that the presence of higher-order single-cavity mechanical modes means that more than one pair of supermodes can be created. Altogether, our simulations indicate that the angle $\theta$ can be used to create a set of even and odd mechanical supermodes that will persist in the presence of fabrication imperfections.

\subsection*{Optical strong coupling with active control}
\begin{figure*}
    \centering
    \includegraphics[width=\textwidth]{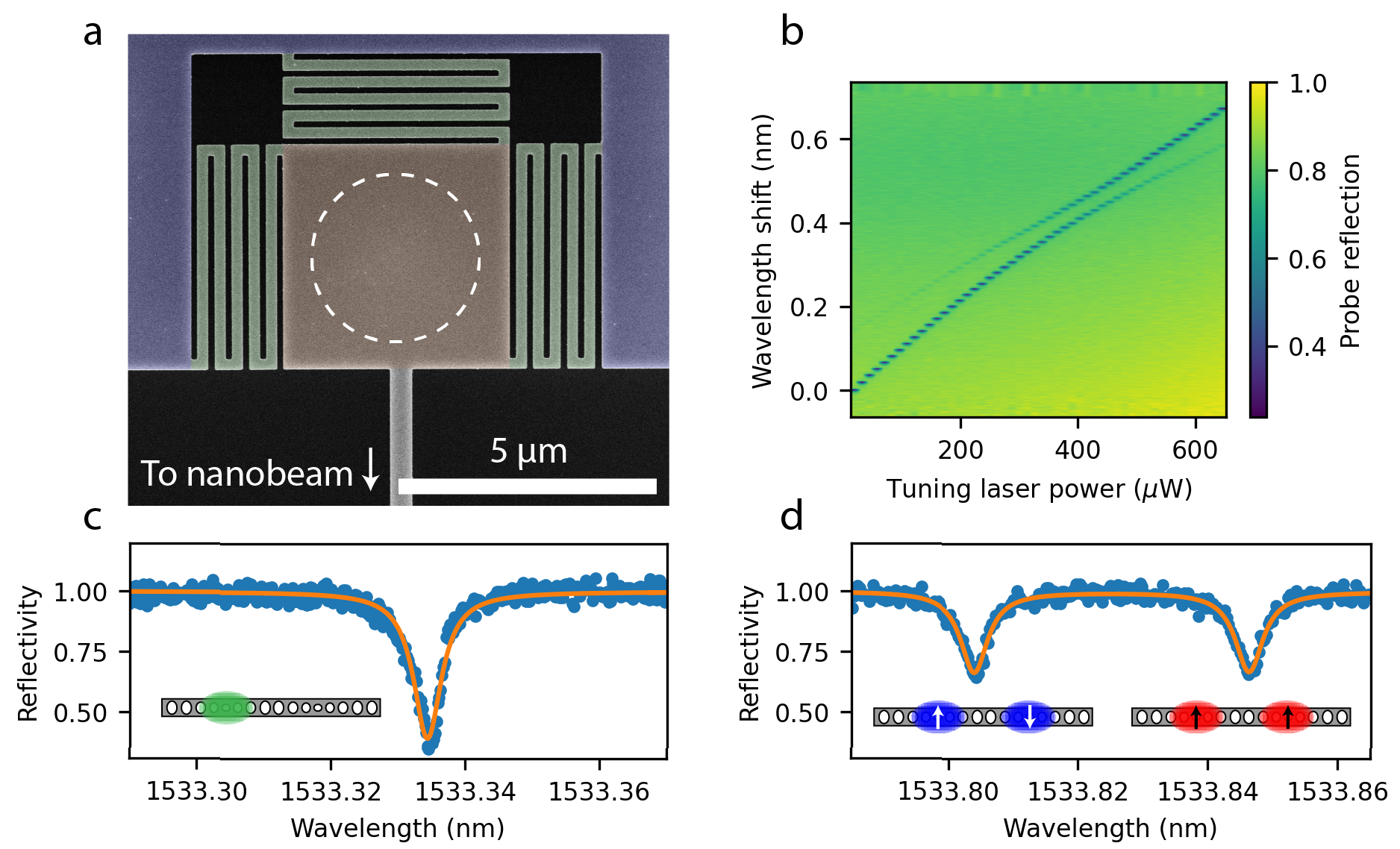}
    \caption{\textbf{Thermal tuning of the optical modes for frequency matching of coupled modes.} a) False-coloured scanning electron microscope image of the thermally-optimised tuning structure between device (bottom) and substrate (top, purple). The meandering tethers (green) limit the flow into the substrate of heat generated by the green laser that is focused on the pad (orange, white circle indication of laser focus scale). b) Spectrogram of IR probe reflection for various wavelengths and power of the tuning laser, showing how a localised mode is tuned to form supermodes at the anticrossing with the other cavity optical mode. c) An example optical reflection trace without thermal tuning. d) Reflection trace with optical modes tuned to the same frequency. Reflectivity traces have been normalised to 1 for off-resonant optical input.}
    \label{fig:opt}
\end{figure*}

Due to fabrication imperfections, the actual optical resonance wavelengths of the left and right modes vary randomly with a typical difference of the order of 1 nm for a design wavelength of 1550 nm. As such a detuning will generally prevent the optical modes from hybridising and precludes enhancement effects, a post-fabrication tuning technique is needed. To allow active tuning, we exploit the temperature dependence of the material refractive index, which in turn controls the resonance wavelength. Creating a variable thermal gradient over the two cavities then allows for control of the inter-cavity detuning \cite{Pan2008}. 

Here, we create a thermal gradient by illuminating the support structure at one end of the nanobeam with a 532 nm green laser spot. We design the support structure (see Fig. \ref{fig:opt}a) to optimise the strength of the achieved temperature gradient. Where the device connects to the support structure, a square pad is thermally isolated from the rest of the sample by thin, meandering tethers. These tethers limit the flow of laser-generated heat into the sample, allowing the suspended device to reach a higher temperature and thus significantly improving the tuning range. See SI for thermal simulations of the support structure.

We characterise the device optical properties by a measurement of reflectivity through one of the waveguides coupled to a single cavity. For an untuned device, the reflectivity typically shows one, localised, optical mode (see Fig. \ref{fig:opt}c), here with $\kappa/(2\pi) = 632$ MHz and $\kappa_\mathrm{ex}/(2\pi) = 120$ MHz. When the tuning laser is applied, the resonant wavelength increases, at a faster rate for the mode closest to the chosen heating pad than for the distant mode. For the correct tuning, this will lead to an anticrossing between the left and right cavity modes (see Fig. \ref{fig:opt}b). At this point, two optical modes ($\kappa_\mathrm{o}/(2\pi) = 642$ MHz, $\kappa_\mathrm{e}/(2\pi) = 653$ MHz) are visible through our interrogation of a single cavity (see Fig. \ref{fig:opt}d), demonstrating the formation of two delocalised supermodes. From the minimal distance between the two supermodes, the inter-cavity coupling can be extracted to be $2J_\mathrm{O} = 5.4$ GHz, a value which is less than one optical linewidth away from the mechanical frequency at around $5$ GHz. This, together with a large sideband resolution factor of $\Omega/\kappa \approx 7.5$, means that our device is capable of enhancing linear and nonlinear optomechanical transduction. 

Importantly, we see no significant broadening of the optical modes with increasing tuning power, indicating that the tuning laser does not induce additional optical absorption and that fluctuations in the tuning are of a size well below the optical linewidth. Note that, with the optical modes tuned, the drop in reflectance on resonance is less pronounced than for a detuned device, because the effective outcoupling of a supermode to a single waveguide is lower than for a localised mode, moving the undercoupled device further away from critical coupling ($\kappa_\mathrm{ex} = \kappa/2$). This could be overcome by changing the designed outcoupling rate of the device accordingly. 

\subsection*{Linear transduction of mechanical supermodes in a coupled-mode system}
We now study the effect of multiple optical modes on the transduction of mechanical motion. We measure the thermomechanical motion at the mechanical frequency $\Omega$, which has an average amplitude that remains constant between different measurements (see Discussion section). In this way, it allows us to compare the strength of the optomechanical transduction of mechanical motion between detuned and tuned systems.

\begin{figure*}
    \centering
    \includegraphics[]{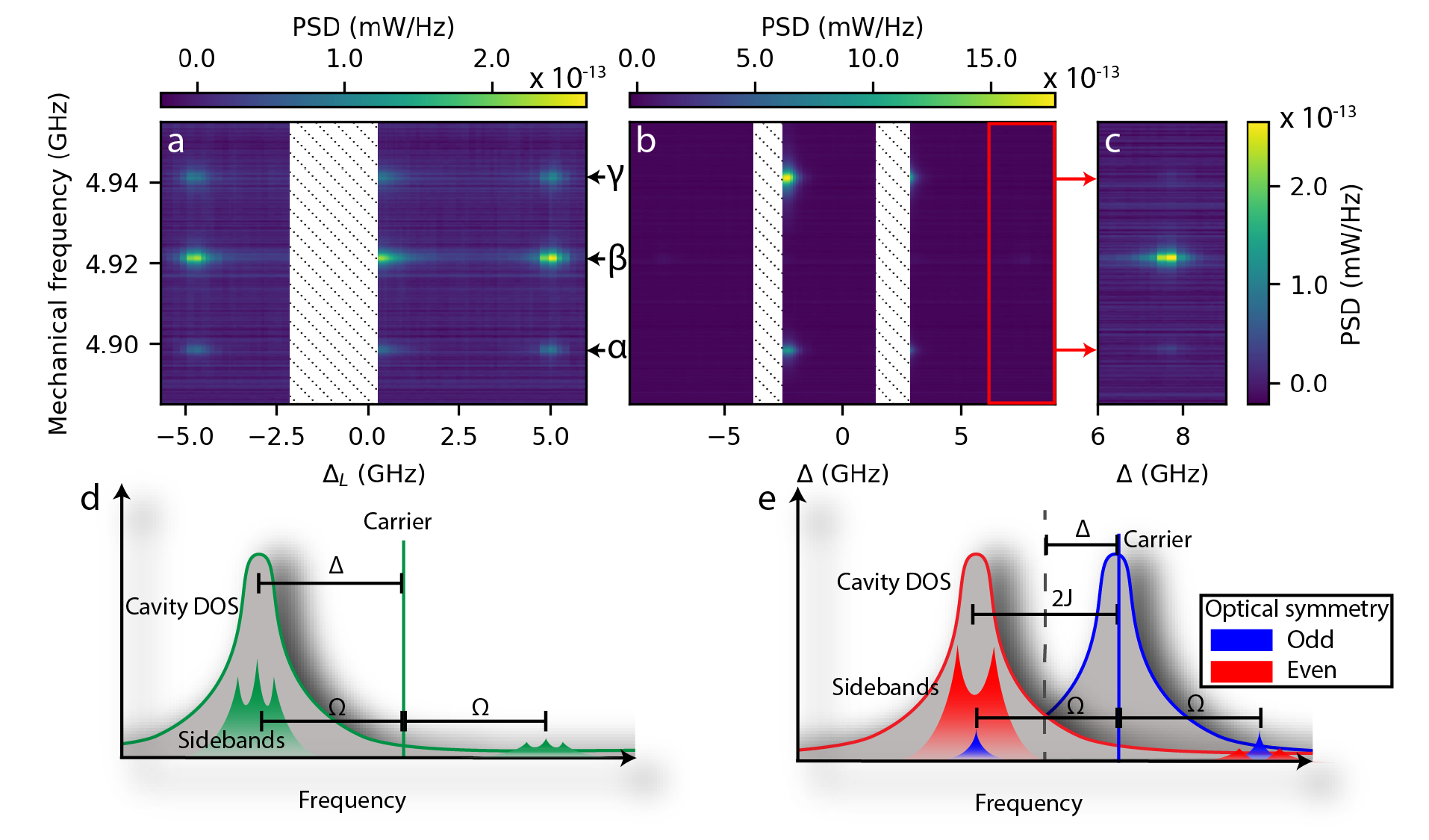}
    \caption{\textbf{Linear transduction of mechanical modes with different symmetries.} Panels a,b,c show spectrograms of photocurrent PSD for various laser-cavity detunings $\Delta$ normalised to 1 \textmu W input power. a) Spectrogram for detuned cavities. b) Spectrogram for a tuned cavity, with c) an additional narrow sweep with more optical power, scaled to match data in b. Hatched regions indicate $\Delta$ values not accessible due to thermal bistability. Optical input powers used: 1.3 \textmu W (a), 1.5 \textmu W (b) and 6.8 \textmu W (c). Panels d and e are diagrams illustrating the different frequency components and resonance conditions. d) Diagram for detuned devices and $\mathrm{Re}(\Delta_\mathrm{L}) = \Omega \approx 4.9$ GHz, showing the sideband resonance condition from single-mode optomechanics. e) Diagram for tuned devices and $\mathrm{Re}(\Delta) = J_\mathrm{O} \approx 2.7$ GHz, showing enhancement per input power for odd mechanical modes. DOS: density of states.}
    \label{fig:EDFA}
\end{figure*}

We use a setup that directly detects intensity fluctuations in the reflected light, using an erbium-doped fiber amplifier (EDFA) and a fast photodiode (See Methods section for details). In Fig. \ref{fig:EDFA}a, we plot the photocurrent power spectral density (PSD) for the thermally detuned system while varying the detuning between the infrared laser and optical mode. We observe three mechanical modes, which we label $\alpha$, $\beta$ and $\gamma$ in order of increasing frequency. For all modes, the transduced signal peaks at $\mathrm{Re}(\Delta_\mathrm{L}) = -\Omega$, $\mathrm{Re}(\Delta_\mathrm{L}) = 0$ and $\mathrm{Re}(\Delta_\mathrm{L}) = \Omega_m$, corresponding to the resonance condition for the upper sideband, laser, and lower sideband, respectively (as per Fig. \ref{fig:EDFA}c). For the hatched area no data could be taken, as for the required input power ($\approx 1$ \textmu W) these values of $\Delta$ are not reachable due to thermal bistability.

By performing these measurements via both the right and left waveguide, interrogating the right and left localised optical modes, we find that these mechanical modes are present in both optical cavities with comparable coupling strengths, showing that the mechanical modes are delocalised supermodes. In a perfect system, these supermodes have either an odd or even symmetry between the left and right halves of the device. We calibrate the vacuum optomechanical coupling rates $g_\mathrm{i,j}$ ($i \in \{\mathrm{L,R}\}$, $j\in \{\alpha,\beta,\gamma\}$) using frequency noise calibration \cite{Gorodetsky2010} (see Methods section for more details). We find that $|g_\mathrm{L,j}|/(2\pi) = (229, 517, 505)$ kHz and $|g_\mathrm{R,j}|/(2\pi) = (375, 608, 443)$ kHz.

Next, in Fig. \ref{fig:EDFA}b, we perform similar measurements for the tuned system. Note that for this configuration, $\mathrm{Re}(\Delta_\mathrm{L})=\mathrm{Re}(\Delta_\mathrm{R})=\mathrm{Re}(\Delta)$ and $\mathrm{Re}(\Delta)=0$ when the measurement laser is exactly between the two optical supermodes. Strikingly, the transduction is now dominated by signal for $\mathrm{Re}(\Delta) = \pm J_\mathrm{O}$, where the laser is on resonance with one of the supermodes. Here, importantly, only modes $\alpha$ and $\gamma$ are visible. Fig. \ref{fig:EDFA}c shows an additional dataset taken around sideband resonance $\mathrm{Re}(\Delta) = J_\mathrm{O} + \Omega$ with more optical power, showing that, conversely, for this detuning, only mode $\beta$ is significantly transduced onto the optical field. Note that there are now two hatched regions, corresponding to the inaccessible red flanks of the two optical supermodes.

The difference in transduction between modes $\alpha$ and $\gamma$ and $\beta$ is determined by the symmetry of the modes. As explained below Eq. \ref{eq:superm_ham} odd mechanical modes create a cross-mode coupling, meaning that they create sidebands in the optical mode with opposite symmetry to that of the carrier light, whereas even mechanical modes create self-mode coupling and thus sidebands with the same symmetry. The data in Fig. \ref{fig:EDFA}b and c suggests that mechanical modes $\alpha$ and $\gamma$ are predominantly odd, whilst mode $\beta$ is even. Fig. \ref{fig:EDFA}d illustrates this argument: the carrier, predominantly exciting the even mode, has sidebands of modes $\alpha$, $\beta$ and $\gamma$ that have the same frequency as the odd optical mode. However, only the sidebands scattered from modes $\alpha$ and $\gamma$ have the odd optical symmetry and are thus resonantly enhanced. The mode $\beta$ sideband is effectively off-resonance. In fact, modes $\alpha$ and $\gamma$ experience the enhancement of linear transduction mentioned in the introduction, which we discuss in more detail in the following.

\subsection*{Enhanced linear transduction in a coupled-mode system}
\begin{figure}
    \centering
    \includegraphics[]{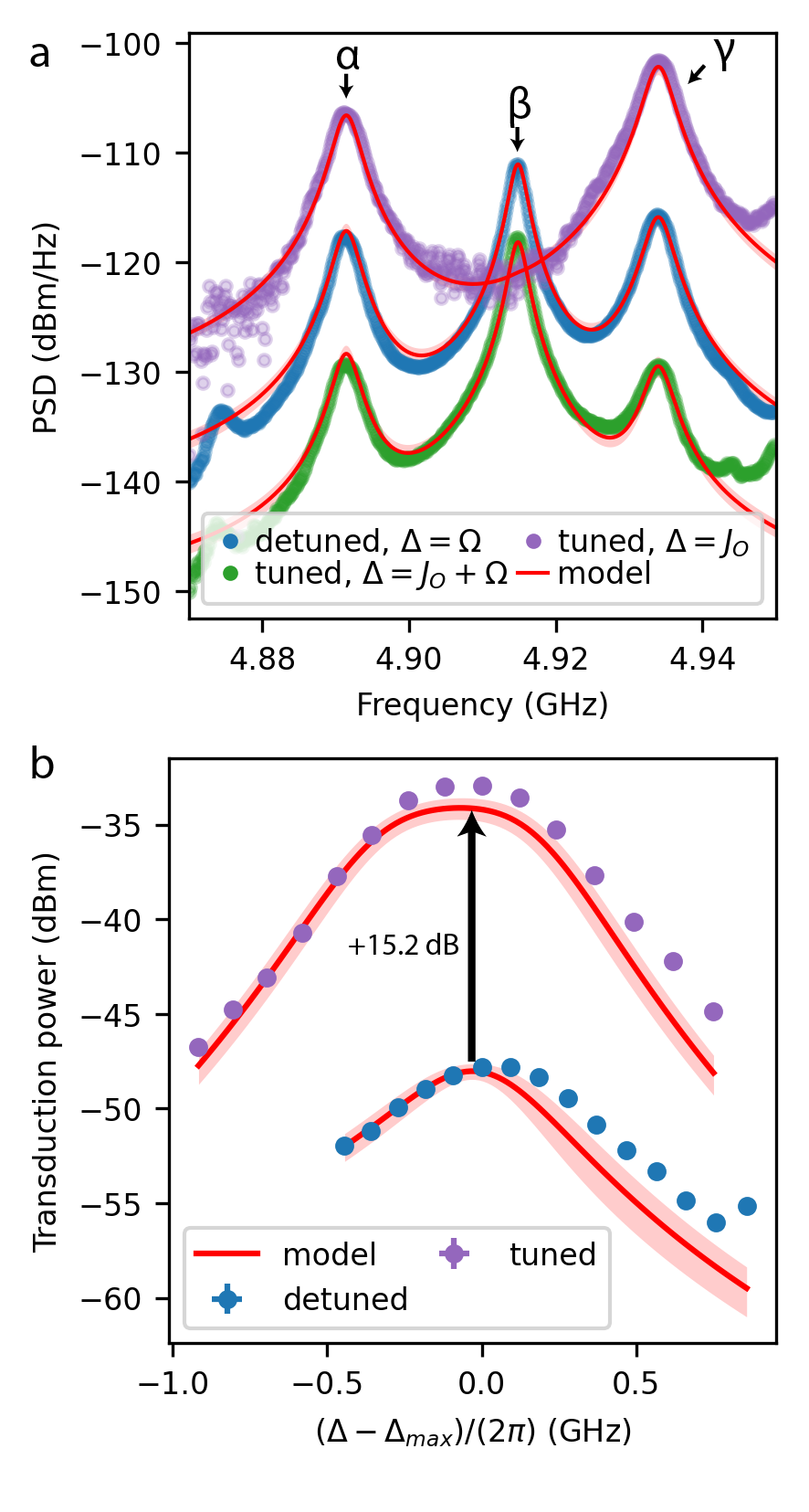}
    \caption{\textbf{Enhanced linear transduction in coupled-mode devices} a) photocurrent PSDs per \textmu W input power for detuned and tuned devices for several interesting detunings $\Delta$, showing enhancement of transduction for tuned devices. Red curve shows the model scaled only to the blue trace, with the highlighted area displaying possible variation due to parameter uncertainty. Input powers used: 36 \textmu W (blue), 68 \textmu W (green), 109 nW (purple), exact values of $\mathrm{Re}(\Delta_\mathrm{max})$: 5.00 GHz (blue), 7.61 GHz (green), 2.45 GHz (purple). b) The fitted area of mode $\gamma$ for detuned and tuned  cavities, varying $\Delta$ around the optimum value. Error bars in the $y$-direction are standard deviations in the fitted area uncertainty. In the $x$-direction they are given by standard deviation of the cavity resonance frequency fit. Both are smaller than the marker size.}
    \label{fig:lin}
\end{figure}

We now quantify the strength of the transduction signal by using heterodyne detection. This allows for better signal-to-noise ratio and for quantitative comparison between linear and nonlinear transduction later on. Using this setup, we perform narrow sweeps around optimum laser-cavity detunings in detuned and tuned systems. For each sweep, the trace with the largest transduction is plotted in Fig. \ref{fig:lin}, normalised to 1 \textmu W of input power. By keeping track of other experimental parameters, direct quantitative comparison between traces is possible (see Methods for more details).

We compare the optimum transduction per input power for a detuned cavity (blue data, $\mathrm{Re}(\Delta_\mathrm{L}) = \Omega$) to that of a tuned cavity (purple data, $\mathrm{Re}(\Delta)=J$) and see a clear enhancement for odd mechanical modes $\alpha$ and $\gamma$, but suppression of even mode $\beta$, just as in Fig. \ref{fig:EDFA}. The enhancement of mode $\gamma$ is stronger than for $\alpha$, which is expected as mode $\alpha$ also has a significant component of even symmetry, as can be seen from the different magnitudes of $g_\mathrm{L,a}$ and $g_\mathrm{R,a}$. We also show a trace for tuned cavities with a different detuning (green data, $\mathrm{Re}(\Delta) = J_\mathrm{O} + \Omega$), which exhibits the opposite effect: a stronger suppression of modes $\alpha$ and $\gamma$ than of mode $\beta$, which can also be understood by comparing the sideband frequencies and symmetries to the optical modes.

These effects can be explained using the theoretical model described above. From Eq. \ref{eq:lintransfulltun} for an odd mechanical mode in a tuned system, it can be seen that transduction can be optimised for $\mathrm{Re}(\Delta)=J_\mathrm{O}$, $J_\mathrm{O}=\Omega/2$. Both terms in the denominator are minimised simultaneously, which can be interpreted as resonance of both the input light and a sideband, as in Fig. \ref{fig:EDFA}e. Such simultaneous resonance is not possible if the mechanical mode has even symmetry or if the optical modes are detuned and thus the coupled-mode system shows transduction that is enhanced with respect to a single mode system. The full equations of our model (see Methods section) are plotted as red lines alongside the data in Fig. \ref{fig:lin}. The model uses as parameters the independently measured optical parameters ($\kappa_i,\kappa_\mathrm{ex,L},J_\mathrm{O}$), OM coupling constants ($g_\mathrm{i,j}$) and the mechanical frequencies and linewidths extracted from a fit of the detuned cavity data with low optical power. The model prediction is scaled overall to the detuned cavity data (blue) to find the unknown photodetector conversion factor, but after that gives completely independent predictions for the tuned cavity transduction. We find excellent agreement between the model and measured data, further strengthening our conclusion about the symmetries of the mechanical modes and the enhancement mechanism at play. The shaded areas around the red lines indicate possible variation in the predicted trend due to uncertainty in the input parameters (see SI for more details).

Finally, we quantify the enhancement of transduction for mode $\gamma$ by fitting the total area of the signal, giving us the total transduced power. In Fig. \ref{fig:lin}b, we plot this power as a function of the laser-cavity detuning around the optimum point for tuned and detuned cavities. We find an enhancement of a factor of 32.9 $\pm$ 0.5 (+15.2 dB) between the optimal transduced powers, where uncertainty is dominated by error on the measurement of the low input power. Note that this enhancement is for constant input power to both the detuned and tuned system. Again, the red lines give the powers extracted from the model, using independent parameters and using the same scaling as in Fig. \ref{fig:lin}a. We compare this enhancement to the ideal value of $\left( \Omega/\kappa \right)^2 \approx 56$. As confirmed by numerical calculations, the lower realised enhancement can be explained well by the non-ideal value of $J_\mathrm{O}$, namely that $J_\mathrm{O}-\Omega/2\approx0.8\kappa$. 

\subsection*{Multimode enhancement of optomechanical nonlinearity}
\begin{figure*}
    \centering
    \includegraphics[]{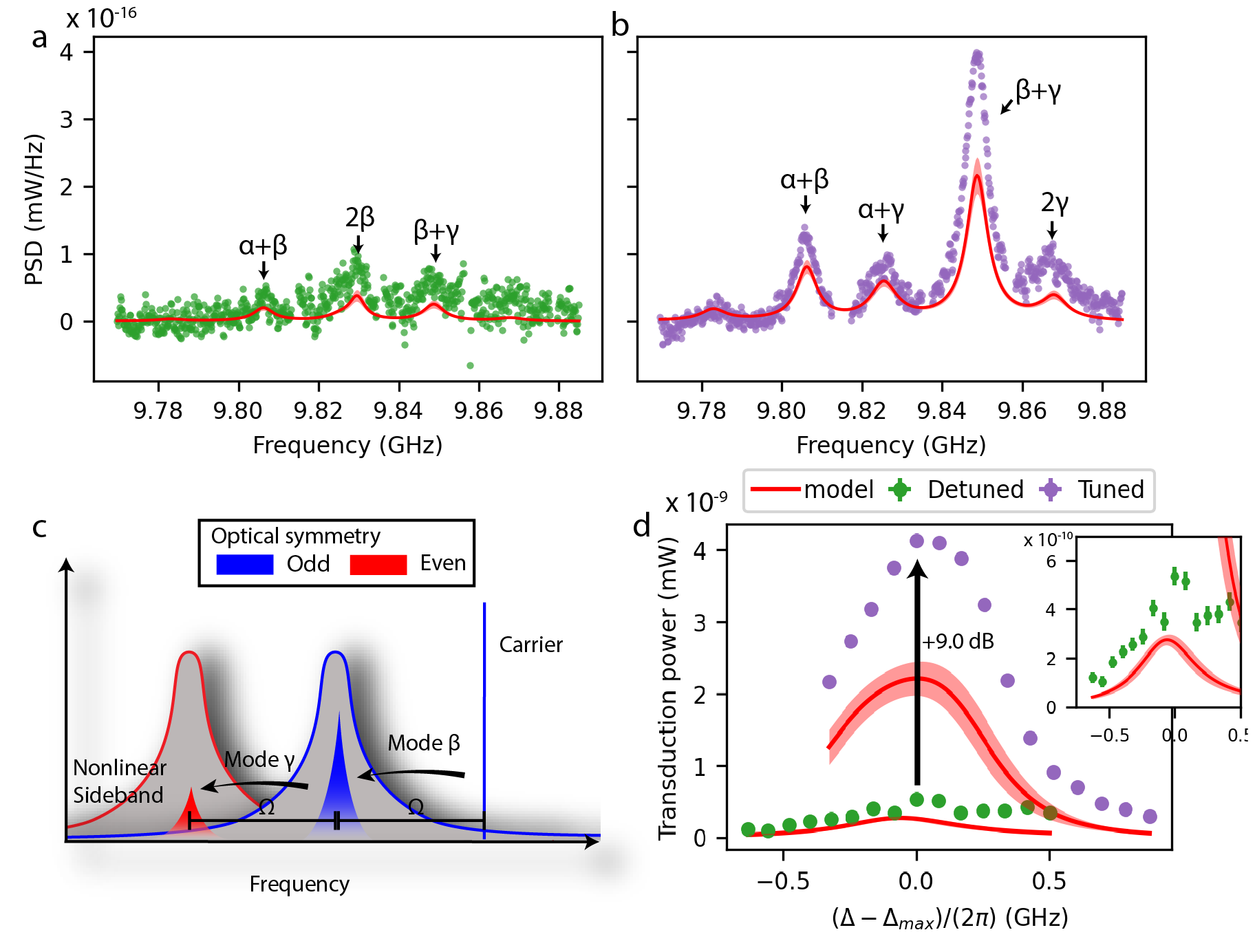}
    \caption{\textbf{Enhanced nonlinear transduction} In panels a and b, we compare the optimal nonlinear signal for detuned (a) to tuned (b) modes. PSD was normalised to per cavity photon. The red line is the independent model prediction, the shaded region the estimated uncertainty on the model. The detuning (power used) was 5.15 GHz (69 \textmu W) and 7.52 GHz (152 \textmu W) for a and b, respectively. c) A schematic representation of the optimal resonance condition for scattering from mode $\beta$ and then mode $\gamma$. d) The fitted area of the $\beta+\gamma$ tone for detuned and tuned modes while varying $\Delta$ around the optimal point. The inset is a zoom-in of the detuned cavity data. Vertical error is fitted area uncertainty, horizontal error is uncertainty in measurement of $\Delta$.}
    \label{fig:nonlin}
\end{figure*}
Nonlinear transduction of mechanical motion is detectable as fluctuations at twice the mechanical frequency. In the weak coupling regime, nonlinear sidebands can be viewed as being created by sequential scattering of light from first-order sidebands \cite{burgwal2020}. This process can involve the same mechanical mode twice, or combine different modes, and results in fluctuations at $\Omega_j + \Omega_k$, where $j,k \in \{\alpha,\beta,\gamma\}$. Classically, a nonlinear sideband contains information about $x_j x_k$, the position-product of mechanical modes $j,k$ creating the sideband. In the quantum regime, a sideband-resolved system does not allow detection of position squared, as can be seen by filling in $\hat{x} = x_\mathrm{zpf}(\hat{b}^\dagger + \hat{b})$ in Eq. \ref{eq:qua_ham}: not all resulting terms will be resonant simultaneously \cite{Thompson2008}. Instead, for on-resonant driving, the optical frequency is determined by the phonon number $\hat{b}^\dagger \hat{b}$, in principle allowing for QND measurement of phonon number. The resulting signal will be centered around zero frequency, making it very hard to detect in our setup. In our experiment, we detect second-order sidebands at $-2\Omega_\mathrm{m}$ away from the carrier, which, in terms of the Hamiltonian, corresponds to the term $\propto \hat{b}^\dagger \hat{b}^\dagger$ that can be used for mechanical squeezing \cite{Nunnenkamp2010} and heralded two-phonon generation \cite{burgwal2020}.

In Fig. \ref{fig:nonlin}a and b, we show the photocurrent spectra of nonlinear transduction in detuned (a) and tuned cavities (b) for optimal detuning ($\mathrm{Re}(\Delta_\mathrm{L}) = \Omega$ and $\mathrm{Re}(\Delta) = J_\mathrm{O} + \Omega$, respectively). The spectra are normalized to one intracavity photon, isolating enhancement of the nonlinear optomechanical processes inside the device from input resonance effects, such as the linear enhancement. Moreover, the intracavity photon number is the limiting factor in many experiments due to heating \cite{riedinger2018,forsch2020,mirhosseini2020}. Normalising to input power instead would reduce the tuned cavity PSD by a factor of roughly 2.

The nonlinear spectra contain several peaks, which can all be attributed to a specific mixing of two mechanical modes by matching the frequency to the sum frequency of two linear transduction peaks (see Fig. \ref{fig:nonlin}a and b). There is a clear enhancement of signal from several nonlinear scattering processes, most notably $\beta+\gamma$,$\alpha + \beta$ and $2\gamma$. For an intuitive understanding of the relative strength of these processes, we have a closer look at the largest peak, $\beta + \gamma$. In Fig. \ref{fig:nonlin}c, we depict schematically how this particular scattering achieves the optimal resonance condition. The carrier light, exciting mostly the odd optical mode, is resonantly scattered into the odd mode through even mechanical mode $\beta$, and subsequently scattered resonantly by odd mode $\gamma$ into the even (opposite symmetry) optical mode. As such, the process is resonant and symmetry-conserving, ensuring maximal enhancement. The $2\gamma$ process is also enhanced and can be described by the simplified transduction expression in Eq. \ref{eq:nonlintransfulltun}. For $\mathrm{Re}(\Delta)=J_\mathrm{O}+\Omega_\gamma$ and $J_\mathrm{O}=\Omega_\gamma/2$, both terms in the denominator are minimized, which can be interpreted as simultaneous resonance of first and second sidebands, and transduction is enhanced over a single-mode device. The $2\gamma$ peak is less strong than the $\beta+\gamma$ peak, as the former requires carrier occupation of the even mode, which is further detuned from the laser.

We compare our experimental results to the model for nonlinear transduction. Note that this model gives an independent prediction, as it is calculated with independently measured system parameters and scaled only once to linear transduction data of the detuned system. In Fig. \ref{fig:nonlin}, we have plotted the model as a red line. We see that the different nonlinear transduction peaks and their relative sizes are captured well by the model. Overall, the model seems to predict less signal than measured experimentally with a difference larger than expected based on statistic uncertainty in measured system parameters. This can possibly be explained by a systematic error in the determination of $g_\mathrm{L(R),j}$ or $\kappa_\mathrm{L(R)}$, parameters that control the ratio of linear to nonlinear transduction. We note indeed that this deviation is present both in the single-cavity and coupled-cavity system.

To quantify the degree of enhancement, we compare the total power of scattering process $\beta + \gamma$ for detuned and tuned cavities. In Fig. \ref{fig:nonlin}d, we plot the fitted areas and the model for a sweep of laser-cavity detuning $\Delta$ around the optimal value. We find an enhancement of a factor 8.0 $\pm$ 0.6 (+9.0 dB), demonstrating enhanced nonlinear optomechanical processes in a coupled-mode device by direct comparison to a single-mode configuration in the same device. The full model, based on the fitted sample parameters, predicts an enhancement at resonant detuning of a factor 8.3, in good agreement with the experimental data obtained by dividing the data of single and coupled-cavity device configurations. We note that the ideal theory for optimally detuned coupling predicted an enhancement of $2\left( \frac{\Omega}{\kappa} \right)^2$ for nonlinear scattering from a single mechanical mode. For this particular scattering $\beta + \gamma$ with two different modes, nonlinear sideband power in the tuned system is reduced by a factor 4, as only scattering from $\beta$ then $\gamma$ is enhanced, while scattering from $\gamma$ then $\beta$ is off-resonant. In the detuned case, both processes have equal amplitude, which means the expected enhancement is $\frac{1}{2}\left( \frac{\Omega}{\kappa} \right)^2 \approx 28$. We have confirmed numerically that this is a good approximation for a system with $2J_\mathrm{O}=\Omega$, and that our lower observed enhancement can be explained through the non-ideal value of $J_\mathrm{O}$.

\section*{Discussion}
We have demonstrated in direct comparison an 8-fold enhanced nonlinear transduction in our coupled mode system, as well as a 33-fold enhanced linear transduction with respect to input power. This demonstration confirms experimentally the idea that optomechanical nonlinearity can be enhanced in a sideband-resolved coupled-mode system. The enhancement was determined by using two configurations of the same device, either tuning the two optical modes to the same frequency, or detuning one completely, effectively removing it from the system and leaving a single-mode device with the same parameters. The mechanical modes remain delocalised, giving multiple mechanical modes even in the single-cavity configuration, at the cost of the vacuum optomechanical coupling $g_0$ being reduced by $\sqrt{2}$ from an uncoupled single cavity, due to the increased mass of the modes.

For the coupled-mode device, we identified optimal enhancement values of optomechanically scattered powers of $\mathcal{E}^\mathrm{lin}=\left(\frac{\Omega}{\kappa}\right)^2$ and $\mathcal{E}^\mathrm{qua} = 2\left(\frac{\Omega}{\kappa}\right)^2$ for linear and nonlinear transduction, respectively. These enhancement factors could, however, both be increased to $\left(2\frac{\Omega}{\kappa} \right)^2$. Tuning the system from effectively single-mode to coupled-mode reduces the effective coupling rate of the optical eigenmodes with the outcoupling waveguide, resulting in less cavity photons and smaller cavity-to-detector efficiency thus giving the lower theoretical maxima we find. This can be overcome by designing the individual cavities with a larger $\kappa_\mathrm{ex,L(R)}$. Next, when comparing experimentally found enhancement to these theoretical values, we find a deviation because of the non-ideal optical coupling rate. A further fine tuning of the optical coupling rate $J_\mathrm{O}$ will allow the device performance to approach optimal enhancement. We also note that some nonlinear scattering processes can be selected by optical excitation of only one particular supermode, which can be achieved by exciting via both on-chip waveguides simultaneously and (anti)symmetrically. Altogether, the maximal nonlinear enhancement of $(2\Omega/\kappa)^2\approx 225$ can be approached in this device by simple redesign within existing possibilities, without the need to further increase optical quality factor. 

In determining the enhancement, we assume implicitly that the amplitude of thermomechanical motion is constant between measurements. This motion is determined by, among other factors, the effective mechanical decay rate and the temperature of the environment. As we use thermal tuning, we do affect the mode environment temperature slightly, although the estimated temperature increase is only 6.4 K (see SI for details), which is marginal compared to the base temperature of 293 K. Therefore, we neglect the effect of this temperature increase in analysis. Next, the effective mechanical decay rate can be changed through the optomechanical interaction via dynamical backaction. In our model, we have included this effect (see Methods section). At the same time, care was taken to keep dynamical backaction effects small during the experiments described here. Still, these effects alter the mechanical position variance slightly, most prominently for mode $\beta$. For the linear enhancement measurement we estimate $\langle x_\beta^2 \rangle_\mathrm{tuned}/\langle x_\beta^2 \rangle_\mathrm{detuned}\approx 0.93$ and for the nonlinear enhancement measurement we estimate $\langle x_\beta^2 \rangle_\mathrm{tuned}/\langle x_\beta^2 \rangle_\mathrm{detuned}\approx 0.96$. The enhancement factors given in the results section have been compensated for this small effect of dynamical backaction.

We have identified the fabrication angle between the device and silicon crystal axis as a degree of freedom to control the mechanical properties of the device, without affecting optical properties. Effectively, additional inter-cavity coupling $J_\mathrm{M}$ was created by leveraging this angle to introduce a new cavity decay channel. Although this, in principle, also increases mechanical radiative decay into the substrate and thus decreases mechanical quality factor, such decrease had only limited effect on our experiment, as these mechanical modes are limited by non-radiative decay channels at room temperature \cite{Alegre2011}. To recover the mechanical quality factor one could terminate nanobeam ends into a structure that has a full phononic bandgap \cite{Chan2012}. Alternatively, further optimisation of coupling region to optimize $J_\mathrm{M}$ and $J_\mathrm{O}$ simultaneously without using anisotropy could be performed. 

Optical post-fabrication tuning was performed through thermal tuning with an auxiliary laser and a thermally isolated device design. The main advantages of this method are the accuracy, reversibility and ease of use. Although temperature increase in the device is only a few Kelvin, the requirement of constant heating can possibly be difficult in cryogenic conditions. Moving forward, the tuning method can be replaced by other methods that are compatible with cryogenics and quantum experiments, such as oxidation tuning \cite{Chen2011,Hennessy2006}, light-induced chemical etching \cite{Gil-Santos2017} or laser-induced gas desorption \cite{kuruma2021}.

Moving forward, the coupled-mode system presented here has several applications, both in the classical and in the quantum regime. First, two optical supermodes at specific frequency separation, combined with several closely spaced mechanical modes provides a very interesting platform for studying mechanical lasing in multiple modes and optomechanical frequency combs \cite{mercade2020,mercade2021}, for which no further device improvements are necessary. In particular, multiple optical resonances allow for the resonant enhancement of specific frequencies from frequency combs, allowing for selective frequency multiplication. For many different applications, our design can be applied to reduce the input power for optical measurement, especially useful in cryogenic applications involving superconducting circuits next to optical components, where optical absorption can degrade performance \cite{mirhosseini2020}. Still in the weak coupling regime, nonclassical states can be generated by heralding, and our device can be used for heralded creation of two-phonon states \cite{burgwal2020}. Moreover, the enhanced nonlinearity could be used to reveal the granularity of mechanical energy by detecting phonon shot noise \cite{Clerk2010}. With further improvements to the vacuum coupling rate $g_0$ and reduction of optical decay rate $\kappa$, approaching the SPSC regime, other quantum applications will come in reach more quickly by use of the coupled-mode device presented here, such as the photon blockade effect for the deterministic generation of single-photon states \cite{Stannigel2012}, measurement-based nonclassical state generation \cite{Brawley2016} and phonon number measurements \cite{ludwig2012}.

\section*{Methods}
\subsection*{Device fabrication}
Devices were fabricated from a silicon-on-insulator wafer with a 220 nm Si device layer on top of a 3 \textmu m SiO\textsubscript{2} sacrificial layer. The Si device layer follows the (100) crystal plane and devices were fabricated at an angle of $\theta=15\degree$ to the $\langle 010 \rangle$ axis. E-beam exposure was used to pattern an HSQ resist layer, followed by development in 25\% TMAH. Anisotropic plasma etching was performed using a mixture of HBr, O\textsubscript{2} and Cl\textsubscript{2} to etch the silicon device layer. Finally, the SiO\textsubscript{2} layer was removed using a 40\% HF etch. After this etch, the device is transported to the setup vacuum chamber within half an hour to prevent oxidation of the Si surface.

\subsection*{Direct detection setup}
The sample is placed in a vacuum chamber which is pumped down and filled with nitrogen back to 0.25 bar to prevent oxidation of the nanobeam surface. Optical connection to the sample was made via a dimpled optical fiber \cite{Hauer2014}. Light from a tunable diode laser (Toptica CTL 1500) was sent into the nanobeam and upon reflection was amplified in an erbium-doped fiber amplifier (EDFA, Calmar Coronado) and detected on a 12 GHz photodiode (New Focus 1544-B). The photocurrent was analysed on a real-time spectrum analyser (Agilent MXA N9020A). To keep track of the optical modes, an additional tunable laser (New Focus TLB-6728) was swept across the optical modes intermittently with low optical power of $\approx 100$ nW. The laser was modulated strongly at 1 MHz and a measure of reflection was obtained with a lock-in measurement of reflected power to overcome detector electronic noise. Determination of optomechanical vacuum coupling rate was done using frequency noise calibration with an electro-optical phase modulator calibrated using a fiber-loop cavity \cite{Gorodetsky2010,Schneider2019}.

\subsection*{Heterodyne setup}
For the heterodyne setup, an additional Toptica CTL tunable diode laser was used as a local oscillator (LO). The two Toptica lasers were locked at a fixed frequency offset by creating sidebands on one laser using an electro-optic modulator and locking the other laser to this sideband. The lock was achieved using a Red Pitaya digital signal processor, applying feedback to the diode current and tuning piezo of the LO laser, and the resulting beating between the two lasers has a linewidth much smaller than the mechanical linewidth. To be able to quantitatively compare different measurements, care was taken to keep constant the LO power, as well as the polarisation overlap between the two lasers and the dimple-to-waveguide coupling efficiency.

\subsection*{Coupled-mode model}
To derive a model that can predict the photocurrent based on parameters of the mechanical and optical modes, we start with the equations-of-motion (EOMs) for the classical optical field amplitudes of left and right optical modes and mechanical mode displacements of modes $\alpha,\beta,\gamma$ in a frame rotating at optical input frequency $\omega_\mathrm{in}$ \cite{Aspelmeyer2014}:
\begin{multline}
    \dot{a}_\mathrm{L(R)} = i\left(\Delta_\mathrm{L(R)} + \sum_{j=a,b,c} g_{\mathrm{L},j} x_j\right) a_\mathrm{L(R)} + iJ_\mathrm{O}a_\mathrm{R(L)} \\ +  \sqrt{\kappa_\mathrm{ex,L(R)}} a_\mathrm{in,L(R)},
\end{multline}
\begin{multline}
    \Ddot{x}_j + \Gamma_j\dot{x} + \Omega_j^2 x = 2\sqrt{\Gamma_j} \Omega_j p_{\mathrm{in},j}\\ + 2\left( g_{\mathrm{L},j}|a_\mathrm{L}|^2 + g_{\mathrm{R},j}|a_\mathrm{R}|^2\right),
\end{multline}
where $\Delta_\mathrm{L(R)} = (\omega_\mathrm{in} - \omega_\mathrm{L(R)}) + i\kappa_\mathrm{L(R)}/2$ contains both detuning between input field and the optical mode frequencies $\omega_\mathrm{L(R)}$ and the optical decay rate $\kappa_\mathrm{L(R)}$. The mechanical modes $j\in\{ \alpha,\beta,\gamma \}$ have frequencies (decay rates) $\Omega_j \;(\Gamma_j)$. Inter-mode optical coupling is given by $J_\mathrm{O}$ and optomechanical coupling is given by $g_{\mathrm{L(R)},j}$. Position is expressed as unitless position $x_j = q_j/x_\mathrm{zpf}$, where $q_j$ is the mode amplitude in meters and $x_\mathrm{zpf}=\sqrt{\frac{\hbar}{2m\Omega}}$ is the zero-point amplitude of the mode, with $m$ the mode effective mass. Finally, optical modes are connected to input fields $a_{\mathrm{in,L(R)}}$ and mechanical modes to thermal bath momenta $p_{\mathrm{in},j}$.

The EOMs are solved in a perturbative fashion \cite{burgwal2020}, $a_\mathrm{L(R)}(t) = \bar{a}_\mathrm{L(R)} + a_\mathrm{L(R)}^\mathrm{(1)}(t) + a_\mathrm{L(R)}^\mathrm{(2)}(t) + ...$, where $\bar{a}_\mathrm{L(R)}$ is the steady-state cavity field and $a_\mathrm{L(R)}^i$ contains all terms of i-th order in $g_{\mathrm{L(R)},j}$. This requires that thermomechanical motion is sufficiently small, i.e. that $g_j \sqrt{n_{\mathrm{th},j}}$, with thermal phonon occupation $n_{\mathrm{th},j} = k_\mathrm{B}T/(\hbar \Omega_j)$, $k_\mathrm{B}$ being the Boltzmann constant and $T$ temperature, is smaller than the optical linewidth $\kappa$. For our system, $g_{\mathrm{L(R)},j} \sqrt{n_{\mathrm{th},j}}/\kappa \approx 0.04$ and we are thus well in the perturbative regime. Also, we assume we connect optically to the left cavity, i.e. $\kappa_\mathrm{ex,R}=0$. 
 
Solving is done in the frequency domain, for which we use the Fourier transform:
\begin{equation}
    A[\omega] = \int_{-\infty}^\infty a(t)e^{i\omega t} dt.
\end{equation}
To transform a product of functions, we use the following identity:
\begin{multline}
    (AB)[\omega] = \int_{-\infty}^\infty b(t)a(t)e^{i\omega t} dt \\ = 1/(2\pi) \int_{-\infty}^\infty A[\omega^\prime] B[\omega-\omega^\prime] d\omega^\prime \\ = 1/(2\pi) A[\omega] \ast B[\omega].
\end{multline}
We find:
\begin{equation}
    \bar{a}_\mathrm{L} = \frac{i\sqrt{\kappa_\mathrm{ex,L}}\Delta_\mathrm{R}}{\Delta_\mathrm{L}\Delta_\mathrm{R}-J_\mathrm{O}^2} \bar{a}_\mathrm{in,L},
\end{equation}
\begin{equation}
    \bar{a}_\mathrm{R} = \frac{-iJ\sqrt{\kappa_\mathrm{ex,R}}}{\Delta_\mathrm{L}\Delta_\mathrm{R}-J_\mathrm{O}^2} \bar{a}_\mathrm{in,L},
\end{equation}
\begin{multline}
    A^\mathrm{(1)}_\mathrm{L(R)}[\omega] = \sum_\mathrm{j=\alpha,\beta,\gamma}\\ \frac{Jg_{\mathrm{R(L)},j}\bar{a}_\mathrm{R(L)} - (\omega + \Delta_\mathrm{R(L)})g_{\mathrm{L(R)},j}\bar{a}_\mathrm{L(R)}}{(\omega + \Delta_\mathrm{R})(\omega + \Delta_\mathrm{L})-J_\mathrm{O}^2} X_j[\omega] \\
    = \sum_\mathrm{j=a,b,c} \Tilde{M}_{\mathrm{L(R)},j} (\omega) X_j[\omega],
\end{multline}
\begin{multline}
    A^\mathrm{(2)}_\mathrm{L}[\omega] = \frac{1}{2\pi} \frac{1}{(\omega + \Delta_\mathrm{R})(\omega + \Delta_\mathrm{L})-J_\mathrm{O}^2} \sum_\mathrm{k=\alpha,\beta,\gamma} \big( \\ Jg_{\mathrm{R},k}(A^\mathrm{(1)}_\mathrm{R}[\omega] \ast X_\mathrm{k}[\omega]) - \\ (\omega + \Delta_\mathrm{R})g_{\mathrm{L},k}(A^\mathrm{(1)}_\mathrm{L}[\omega] \ast X_\mathrm{k}[\omega])\big).
\end{multline}
We are interested in the PSD of photocurrent $I$, which is given by \cite{Bowen2016}:
\begin{equation}
    S_\mathrm{II}[\omega] = \frac{1}{2\pi}\int_{-\infty}^\infty \langle I[\omega] I[\omega^\prime] \rangle d\omega^\prime.
\end{equation}
The photocurrent is equal to the optical power (removing the proportionality constant), which for the heterodyne detection is given by:
\begin{equation}
    I[\omega] = \sqrt{n_\mathrm{het}}\sqrt{\kappa_\mathrm{ex,L}}\left( A_\mathrm{out}^\mathrm{(i)}[\omega_-] + (A_\mathrm{out}^\mathrm{(i)}[-\omega_+])^\ast \right),
\end{equation}
with $\omega_- = \omega - \omega_\mathrm{het}$ and $\omega_+ = \omega + \omega_\mathrm{het}$, $\omega_\mathrm{het}$ the heterodyne frequency and $n_\mathrm{het}$ the amount of photons in the LO. The cavity reflected light for non-zero $\omega$ is given by the input-output relation $A^\mathrm{(i)}_\mathrm{out} = \sqrt{\kappa_\mathrm{ex,L}}A^\mathrm{(i)}_\mathrm{L}$.

Now, we need to specify the motion of $X_j[\omega]$. Following Bowen and Milburn \cite{Bowen2016}, we define this to be:
\begin{equation}
    X_j[\omega] = \chi_j(\omega) P_{\mathrm{in},j}[\omega],
\end{equation}
where $\chi_j(\omega)$ is the susceptibility of mechanical mode j, with first-order dynamical back-action correction:
\begin{multline}
    \chi_j(\omega) = 2\sqrt{\Gamma_j}\Omega_j\Big[\Omega_j^2 - \omega^2 - i\omega \Gamma_j -2\Omega_j \big(\\g_{\mathrm{L},j}\bar{a}_\mathrm{L}\Tilde{M}_{\mathrm{L},j}^\ast(-\omega) + g_{\mathrm{L},j}\bar{a}^\ast_\mathrm{L}\Tilde{M}_{\mathrm{L},j}(\omega) \\ g_{\mathrm{R},j}\bar{a}_\mathrm{R}\Tilde{M}_{\mathrm{R},j}^\ast(-\omega) + g_{\mathrm{R},j}\bar{a}^\ast_\mathrm{R}\Tilde{M}_{\mathrm{R},j}(\omega) \big)\Big]^{-1},
\end{multline}
and $P_\mathrm{in,j}$ the thermal bath forcing (momentum) term, which is a white noise with correlation function:
\begin{equation}
    \langle P_{\mathrm{in},j}[\omega] P_{\mathrm{in},k}[\omega^\prime]\rangle = 2\pi n_\mathrm{th}\delta_{j,k} \delta(\omega + \omega^\prime).
\end{equation}
In the calculation of second-order PSD, the correlation function of a product of four thermal bath momenta has to be calculated. To evaluate this, we use the fact that, in thermal equilibrium, the momenta are normally distributed to employ the Isserlis-Wick theorem \cite{Brawley2016}.

Combining all of the previous steps, we can write down expressions for the first and second-order components of $S_\mathrm{II}$:
\begin{multline}
    S_\mathrm{II}^\mathrm{(1)}[\omega] = n_\mathrm{het}\kappa_\mathrm{ex,L} n_\mathrm{th} \sum_j \bigg[ M_j(\omega_-)M_j(-\omega_-) \\ + M_j(\omega_-)(M_j(\omega_-))^\ast 
    (M_j(-\omega_+))^\ast M_j(-\omega_+) \\ + (M_j(-\omega_+))^\ast (M_j(\omega_+))^\ast\bigg],
\end{multline}
with
\begin{equation}
    M_j(\omega) = M_{\mathrm{L},j}(\omega) = \Tilde{M}_{\mathrm{L},j}(\omega) \chi_j(\omega).
\end{equation}

Note that $M_\mathrm{R,j}$ is obtained from $M_\mathrm{L,j}$ by swapping subscripts R and L. For nonlinear transduction, we find:
\begin{multline}
    S_\mathrm{II}^\mathrm{(2)}[\omega] = \frac{n_\mathrm{het}\kappa_\mathrm{ex,L} n_\mathrm{th}^2}{2\pi} \int d\omega^\prime \sum_\mathrm{j,k=a,b,c} \bigg( N_{j,k}(\omega_-,\omega^\prime)\big[ \\ N_{j,k}(-\omega_-,-\omega^\prime) + N_{k,j}(-\omega_-,\omega^\prime - \omega_-) \\ + N_{j,k}^\ast(\omega_-,\omega^\prime) + N_{k,j}^\ast(\omega_-,\omega_- - \omega^\prime) \big] \\ + N_{j,k}^\ast(-\omega_+,\omega^\prime)\big[ N_{j,k}(-\omega_+,\omega^\prime) \\ + N_{k,j}(-\omega_+,-\omega_+-\omega^\prime) + N_{j,k}^\ast(\omega_+,-\omega^\prime) \\ + N_{k,j}^\ast(\omega_+,\omega^\prime + \omega_+)\big] \bigg),
\end{multline}
with
\begin{multline}
    N_j(\omega,\omega^\prime) = \frac{1}{(\omega + \Delta_\mathrm{R})(\omega + \Delta_\mathrm{L})-J_\mathrm{O}^2} \big(\\Jg_{\mathrm{R},k}M_{\mathrm{R},j}(\omega-\omega^\prime)- \\(\omega + \Delta_\mathrm{R})g_{\mathrm{L},k}M_{\mathrm{L},j}(\omega-\omega^\prime)\big) \chi_\mathrm{k}(\omega^\prime).
\end{multline}

\section*{Acknowledgements}
We would like to thank P. Busi, A. Fiore, M. Lodde and P. Neveu for valueable discussions. We thank F. Koenderink for critical reading of the manuscript. This work is part of the research programme of the Netherlands Organisation for Scientific Research (NWO). E.V. acknowledges support from NWO Vidi, Projectruimte, and Vrij Programma (Grant No.680.92.18.04) Grants, and the European Research Council (ERC Starting Grant No. 759644-TOPP).

\section*{Author contributions}
R.B. and E.V. conceived the project. R.B. designed and fabricated the sample. R.B. and E.V. designed the experiment, R.B. built the setup, performed measurements, performed data analysis and derived model equations. E.V. supervised the project. Both authors participated in the writing of the manuscript.

\bibliography{library}

\setcounter{figure}{0}
\setcounter{equation}{0}
\makeatletter 
\renewcommand{\thefigure}{S\@arabic\c@figure} 
\renewcommand{\theHfigure}{Supplement.\thefigure} 
\renewcommand{\theHequation}{Supplement.\theequation}
\makeatother

\clearpage
\section*{Supplementary information}
\subsection*{Experimental setup}
\begin{figure*}
    \centering
    \includegraphics{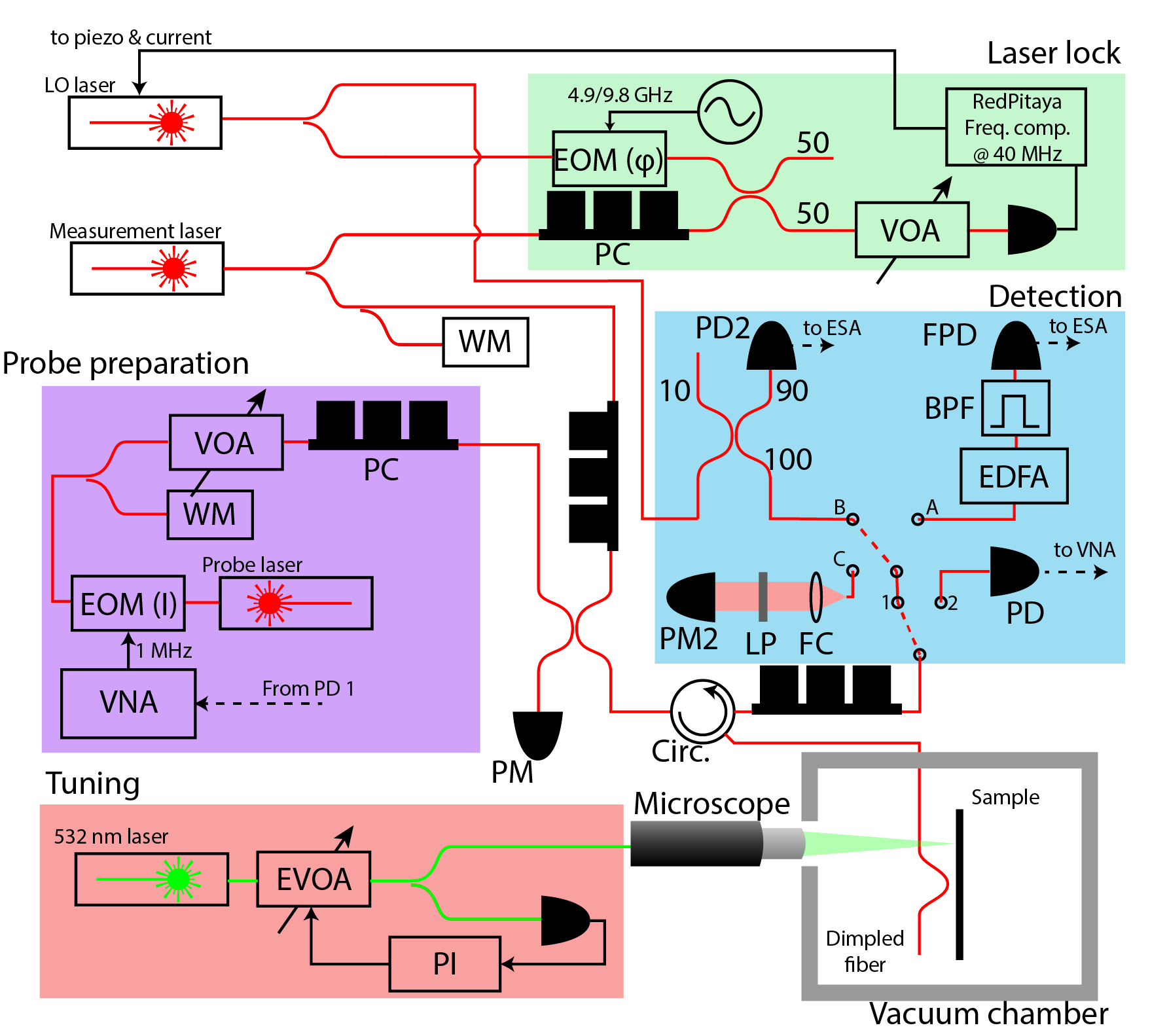}
    \caption{\textbf{A detailed schematic of the experimental setup.} Different coloured areas group the components by function. Abbreviations used: EOM($\phi$/I): electro-optical phase/intensity modulator, (E)VOA: (electronic) variable optical attenuator, WM: wavemeter, ESA: electronic spectrum analyser, VNA: vector network analyser, PI: proportional-integral controller, PD: photodiode, FPD: fast photodiode, PM: powermeter, PC: polarisation controller, FC: fiber-to-freespace coupler, LP: freespace linear polariser, EFDA: erbium-doped fiber amplifier, BPF: band-pass filter}
    \label{fig:setup}
\end{figure*}

Fig. \ref{fig:setup} shows a schematic of the setup used, where different functional parts have been identified with a coloured background.

The laser lock serves to lock the LO and measurement laser at a fixed frequency offset. This is achieved by creating sidebands at $f_\mathrm{lock}^\prime$ of the LO laser and measuring the slow beating of one of these sidebands with the measurement laser that is placed close in frequency. This signal is processed by a Red Pitaya digital signal processor, which uses a digital phase-frequency comparator to stabilize the beating to 40 MHz. To set up the correct signal processing on the Red Pitaya, the PyRPL software package was used \cite{neuhaus2017_SI}. The error signal is fed back proportionally to the LO laser laser current and, after integration, to the piezo of the tunable laser. In this way, the LO and measurement laser at locked at a frequency separation of $f_\mathrm{lock} = f_\mathrm{lock}^\prime + 40 \; \mathrm{MHz}$. By direct measurement it was established that the beating frequency was stabilized to a linewidth much smaller than the mechanical linewidth of $\approx$ 3 MHz.

The probe preparation unit creates an intensity modulation at 1 MHz of the weak probe laser that allows a lock-in measurement of the reflected power of this laser. The probe laser power was kept low enough such that the probe reflection around resonance could be well described by a Lorentzian, with negligible thermal bistability, with a typical power of 100 nW. When sweeping the probe laser, the laser wavelength is recorded on a wavemeter (Bristol Instruments 871) with repeatability better than 0.1 pm. By fitting the lock-in reflection measurement, the cavity resonance wavelength is determined, from which the detuning $\Delta$ can be calculated. 

In the detection block, there are two consecutive switches. In the first switch, option 1 leads to detection of transduction, while option 2 is used for detection of the optical properties of the device by measuring probe beam reflection. Such a measurement of optical properties is done before and after each transduction measurement. In the transduction path, switch option A connects to the direct detection configuration, where light is amplified in an EDFA, filtered at the carrier wavelength by a band-pass filter of 0.8 nm FWHM and detected on a fast photodiode (12 GHz bandwidth). In switch path B, the reflected light is mixed with the local oscillator (LO) and captured on a slower, low-noise photodiode. Switch path C is used before and after each heterodyne measurement for the polarisation check described below.

The vacuum chamber that keeps the sample is filled with nitrogen gas to a pressure of 0.25 bar to prevent oxidation or other contamination of the sample. Some sample degradation is, however, still observed. Over the course of weeks after fabrication, the sample optical mode is observed to shift to higher wavelength and the optical linewidth increases. Additionally, a small shift in mechanical resonance frequencies is observed. This is thought to be a sign of the deposition of some contamination on the device while in the vacuum chamber. The sample can be restored to its original state through a brief oxygen plasma clean, after which device optical and mechanical properties are found to return to original values. The sample was plasma cleaned right before measurements reported in this paper. There is an additional round of cleaning between the data in figures 2 and 3 and figures 4 and 5. Importantly, the data of figures 4 and 5 is taken within one day shortly after plasma cleaning.

Finally, the tuning block serves to digitally control the green laser power and to stabilize it. The green light passes through an electronically-controlled variable attenuator (EVOA), after which a part of the laser power is split off and detected on a photodiode. The detector voltage passed through a proportional-integral controller and fed back to the EVOA.

\subsection*{Tracking relevant setup parameters}
To keep constant all relevant parameters between different measurements, we record input power, reflected power and polarisation. Typically, overall reflectivity, defined as PD reflected power divided by PM input power, is about 13\%. This encompasses losses in the circulator, fiber tapering region and fiber-to-waveguide coupling efficiency, which are all occurring twice, once in the input path and once in the reflection path. The transmission through the dimple when detached from the sample is 50\%. We can estimate a minimum fiber-to-waveguide coupling efficiency of $\sqrt{0.13}\cdot 100\% = 36\%$.

To control the polarisation, the combined LO and reflected light are outcoupled into free space and sent through a linear polariser onto a photodetector. The polarisation of both is optimised for transmission through the polariser, ensuring overlap of polarisation of both modes, and measured powers are recorded. This procedure is repeated after each measurement to make sure no significant drift occurred during measurement.

\subsection*{Further comments on data processing}
The values of laser-cavity detuning $\Delta$ are determined from the cavity frequency, found by fitting the probe reflection sweep, and the laser frequency, determined by the wavemeter. It is expected that the fit uncertainty dominates the noise on detuning, while the actual cavity frequency is more stable. Therefore, for heterodyne measurements that sweep only a small range of detunings, we fit a linear function to the cavity frequency versus laser frequency data to extract smoothed values of detuning. The linear slope can capture small cavity frequency shifts due to laser heating. 

\begin{figure*}
    \centering
    \includegraphics{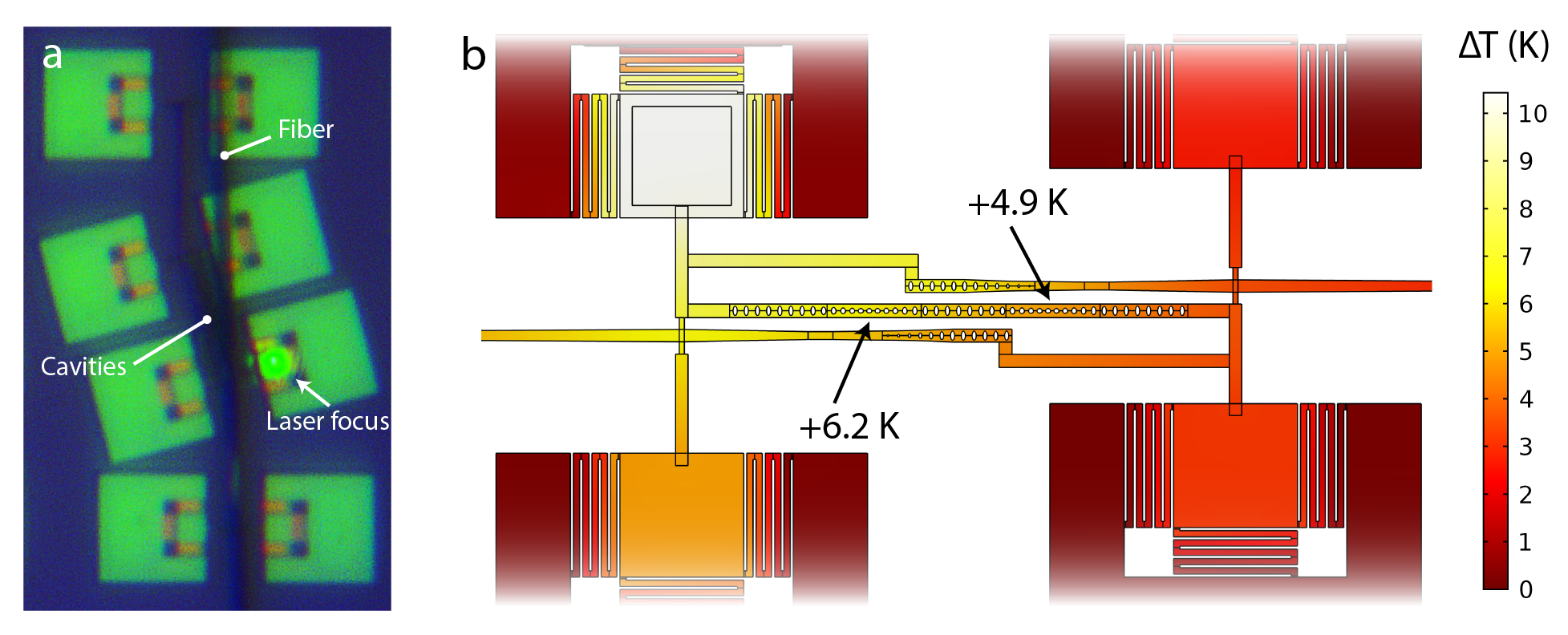}
    \caption{\textbf{Thermal tuning with a green laser heat source} a) An optical microscope image of a fabricated device in the setup, with the 532 nm laser focused on a heating pad and the fiber connected to one of the device waveguides. b) Simulated temperature increase on the device created by a 20 \textmu W dissipation in one of the heating pads, corresponding to 200 \textmu W tuning laser power. Temperature increase in closest and furthest cavity indicated in figure, giving a relative tuning of 1.3 K.}
    \label{fig:SItuning}
\end{figure*}
To isolate optomechanical transduction from the noise background on the heterodyne photodiode, a background subtraction is required. To this end, a noise trace with only the LO is taken before every measurement. The shot noise created by the LO is scaled slightly to match the heterodyne measurement in a frequency region without optomechanical features, after which it is subtracted from the measurement.

\subsection*{Thermal tuning simulation and evaluation}
A 532 nm laser, focused on one of the design heat pads, acts as a heat source for thermal tuning. In Fig. \ref{fig:SItuning}a, we show an optical microscope image of thermal tuning in the setup. We estimate absorbed power to be rougly 10\% of optical power, based on a silicon-air interface reflectivity of 0.3 and an absorption length in silicon for 532 nm of 1.5 \textmu W \cite{wang2013_SI}. 

The thermal design of the heating pads and tethers was evaluated using COMSOL Multiphysics. In Fig. \ref{fig:SItuning}b, we show a stationary thermal simulation in which a heat source is placed in one heating pad and the substrate beneath the device acts as a room-temperature heat sink. Because of the thermally isolating properties of the tethers, the closest cavity reaches +6.2 K, with the temperature difference between the cavities at 1.3 K, for a dissipated power of only 20 \textmu W, corresponding to 200 \textmu W laser power. Importantly, for a design without tethers and thus a more direct connection to substrate, a temperature difference between the cavities of only 0.07 K was achieved, meaning that the use of tethers allows us to create a roughly 20 times stronger temperature gradient across the cavities with the same heating power.

Next, the cavity wavelength shift per Kelvin temperature increase can be calculated using a perturbation theory approach \cite{chan2012a_SI}:
\begin{equation}
    \Delta \lambda = 7.5066\cdot 10^{-2} \lambda \, n_\mathrm{Si}(T)\, \frac{dn_\mathrm{Si}}{dT}, 
\end{equation}
where the unitless numerical coefficient characterises the effect of refractive index on wavelength, $n_\mathrm{Si}$ is the refractive index of silicon and $\lambda$ is the wavelength. Using silicon material properties from \cite{frey2006_SI}, we evaluate that:
\begin{equation}
    \Delta \lambda = 73\;\mathrm{pm/K}.
\end{equation}
Using this value, we can calculate that we expect a absolute wavelength shift of the optical mode closest to the pad of 2.2 pm/\textmu W, and a relative inter-mode shift of 0.47 pm/\textmu W. 

From the experimental data, we can determine the tuning coefficient by applying a linear fit to the data of figure 2b from the main text. For this, we restrict ourselves to powers below 200 \textmu W, away from the anticrossing, where the mode wavelength scales approximately linear with input power. We find a coefficient of 1.15 pm/\textmu W, a little over half of our simulated coefficient. The discrepancy can possibly be caused by the pad and the laser spot not fully overlapping or by our rough estimation of absorbed power. We can also conclude that at perfect mode tuning for about 400 \textmu W optical power, the device temperature is raised by only 6.3 Kelvin. 

\subsection*{Simulation of device mechanical modes}
\begin{figure*}
    \centering
    \includegraphics[width=0.8\textwidth]{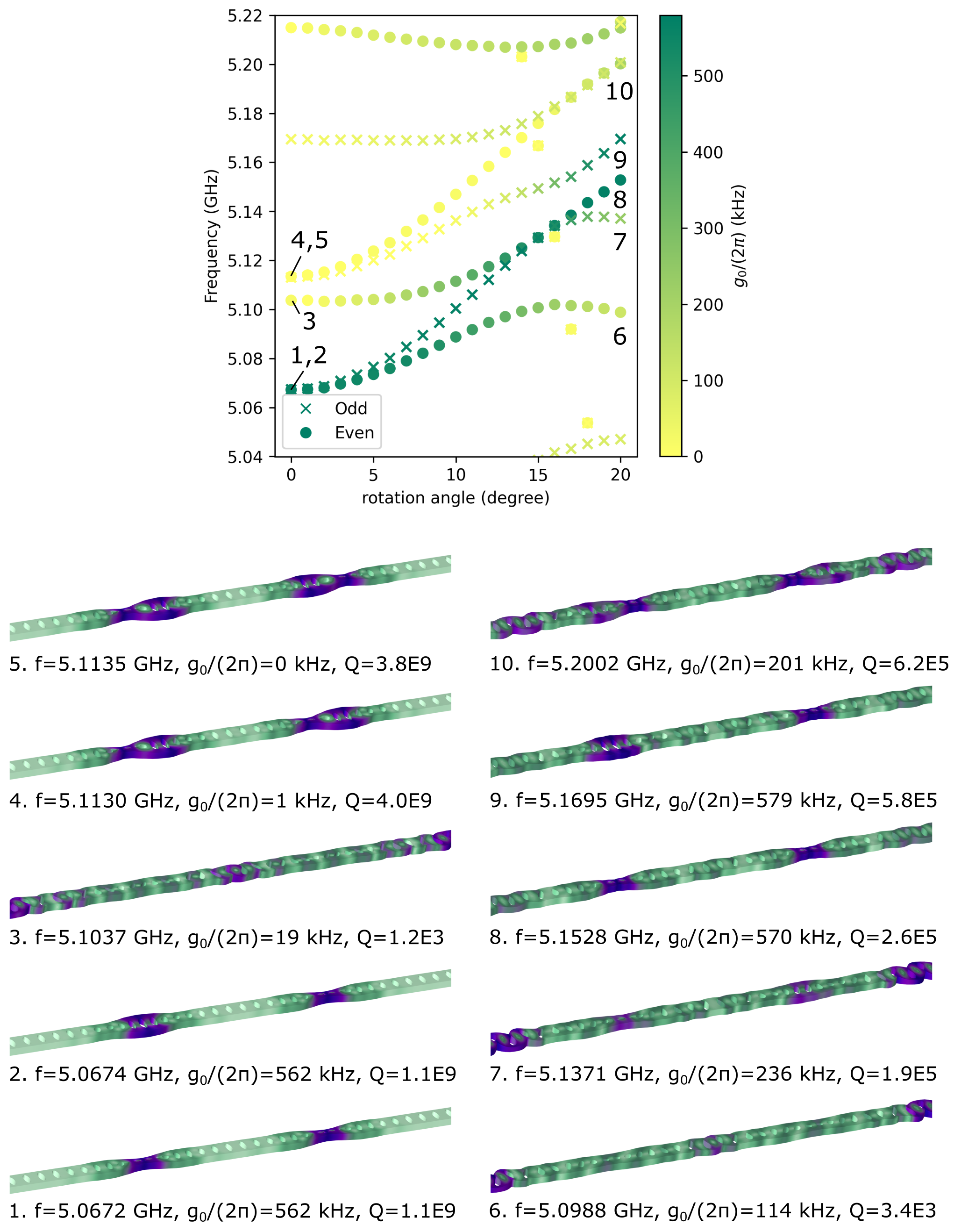}
    \caption{\textbf{Multiple mechanical modes with optomechanical coupling varying with angle to silicon crystal axis.} The plot displays the eigenmodes of the double cavity nanobeam, with the marker indicating the $x$-symmetry of the mode (cross is odd, circle is even). The color of the marker indicates the optomechanical coupling strength $g_0/(2\pi)$. Numbers 1-10 indicate specific eigenmodes at $\theta=0,20$, of which the mode shape is plotted in the panels below.}
    \label{fig:mechmodes}
\end{figure*}
After fabrication, the realised dimensions of the device were measured using a scanning electron microscope (SEM), from which it became clear that the beam width was slightly smaller (average $-11$ nm) and the hole diameter slightly bigger (average $+5$ nm) than the design. Device simulations were performed in COMSOL Multiphysics using these corrected dimensions. The device phononic crystal with two cavities is simulated, terminated at both ends in a bare waveguide with no holes which leads into a perfectly matched layer (PML) to simulate leaking into the substrate. The anisotropic mechanical properties of the crystalline silicon mean that relation between stress and strain in the material depends on the direction of the stress and strain with respect to the crystal axis. This is caught in an elasticity matrix \cite{chan2012a_SI}, which is chosen such that silicon crystal axes $\langle 100 \rangle$ align with the $x$ ,$y$ and $z$ direction of the simulation geometry. The nanobeam geometry is aligned along the $x$ axis and subsequently rotated in plane with an angle $\theta$ to simulate the varying mechanical properties of devices fabricated at different angles to the substate crystal axis.

Fig. \ref{fig:mechmodes} shows how the mechanical eigenmodes of the system change as the angle $\theta$ to the crystal axis is varied. At $\theta=0$, for which the cavities are designed, we see odd and even supermodes of the fundamental cavity modes (panels 1 and 2) and the second-order cavity modes (panels 4 and 5). The mechanical coupling rates $J_\mathrm{M}$ are 2 MHz and 5 MHz, respectively. Of these two only the fundamental modes couple to the optical mode, with coupling rate $g_0/(2\pi)$ reduced by $\sqrt{2}$ with respect to individual cavities due to the higher effective mass. Between the two pairs of modes lies a mode with odd $y$-symmetry (panel 3), not confined by the cavities and thus leaking into the environment more easily, although still confined in the phononic crystal.

When we increase $\theta$, mode frequencies start to shift. The non-zero angle breaks the system's $y$-symmetry and couples modes of different $y$-symmetry. As can be seen in Fig. \ref{fig:mechmodes} panels 6-10 for $\theta=20\degree$, the new eigenmodes appear to be combinations of the $y$-even cavity modes and $y$-odd waveguide mode at $\theta=0$. As such, they are less confined in the cavities and new eigenmodes are at a larger frequency separation, without affecting the optical properties of the device. As a consequence, the quality factors of these modes are lower. However, optomechanical coupling $g_0/(2\pi)$ is not reduced.

\subsection*{Simulation of device optical modes}
\begin{figure*}
    \centering
    \includegraphics[width=0.8\textwidth]{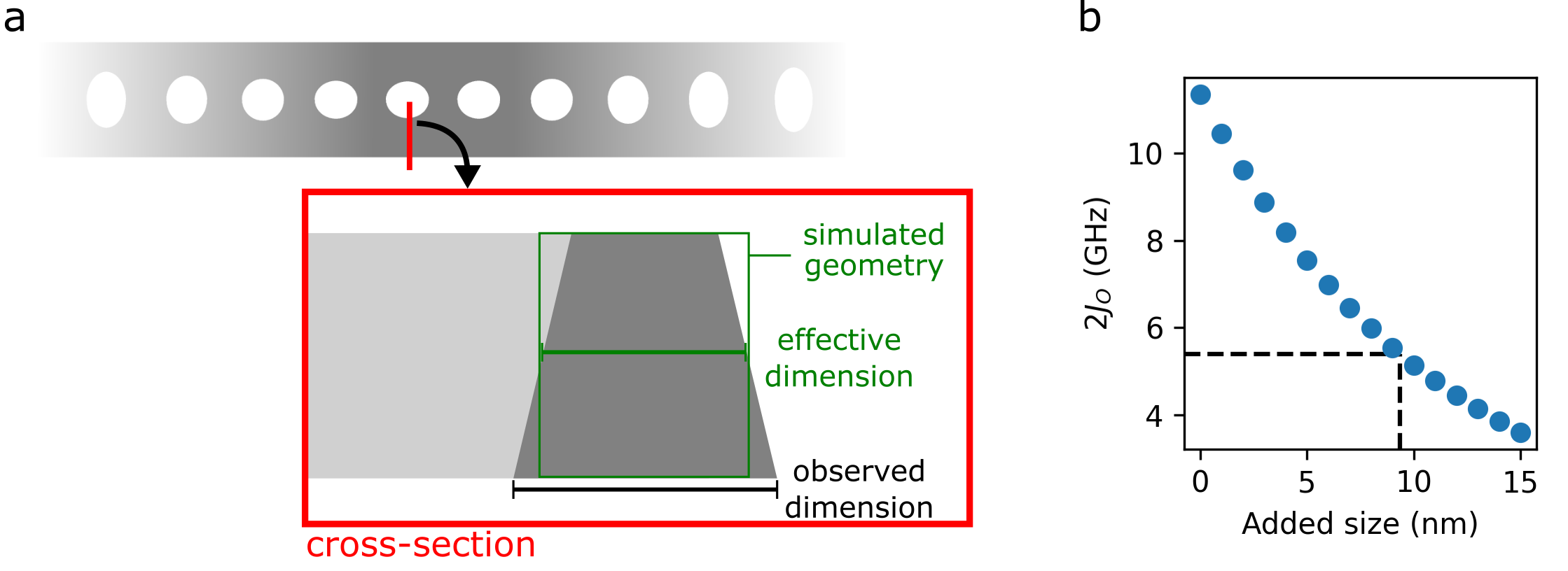}
    \caption{\textbf{Optical coupling $J_\mathrm{O}$ varying due to slanted device sidewalls} a) A schematic display of a cross-cut of a fabricated device which has sidewalls at an angle of $\phi$ from vertical (grey), for which the observed dimension will be an overestimation of device volume. The green outline indicates the geometry that is simulated to estimate the effect of this overestimation. b) Simulated supermode splitting $2J_\mathrm{O}$ as a function of the added hole size and reduced nanobeam width.}
    \label{fig:optmodes}
\end{figure*}
Optical simulations of the device with realistic dimensions give a frequency difference $2J_\mathrm{O}$ between the two supermodes of 11.3 GHz, significantly larger than the 5.4 GHz observed in experiment. Although SEM imaging tells us the in-plane dimensions of the device, it is much harder to accurately determine variation along the vertical direction. For example, slanted sidewalls (see Fig. \ref{fig:optmodes}a) would result in an observed device dimension that overestimates the average dimension along the vertical direction. To emulate slanted sidewalls, we simulate our device with a slightly increased hole size and reduced beam width. In Fig. \ref{fig:optmodes}b, we see that an additional 9 nm, corresponding to a sidewall angle of just $\phi = 4.8\degree$, could explain the observed optical splitting.

The enhancement demonstrated here relies on a carefully tuned inter-cavity optical coupling $J_\mathrm{O}$, and the above simulation suggests that the optical coupling is sensitive to small changes in sidewall angle or hole dimension. Such sensitivity of $J_\mathrm{O}$ to fabrication imperfection can possibly be avoided by creating optical coupling in a zipper cavity configuration \cite{eichenfield2009_SI}, which can be combined with capacitative tuning of the gap for control over $J_\mathrm{O}$ \cite{paraiso2015_SI}. Although this will decouple mechanical modes, enhancement of linear and nonlinear transduction is still possible.

\subsection*{Estimating enhancement factors from coupled-mode model}
Starting from our coupled-mode model equations, we apply some simplifications to extract simple expressions for the approximate enhancement of linear and nonlinear transduction in our particular system.

We start with linear transduction, $S_\mathrm{II}^{(1)}[\omega]$. We assume tuned cavities ($\Delta_\mathrm{L}=\Delta_\mathrm{R}=\Delta$), close to optimal detuning $\Re(\Delta) \approx \Omega_m$ and optimal $J_\mathrm{O}\approx\Omega/2$. We assume that we excite only the odd optical mode, such that $\bar{a}_\mathrm{L} = -\bar{a}_\mathrm{R}=1/\sqrt{2}\bar{a}_\mathrm{o}$, the odd mode steady state being $\bar{a}_\mathrm{o}=\frac{i}{\sqrt{2}}\frac{\sqrt{\kappa_\mathrm{ex,L}}a_\mathrm{in,L}}{\Delta-J}$. Finally, we consider only one odd mechanical mode with $g_\mathrm{R}=-g_\mathrm{L}=g$. Under these conditions, $M_j(\omega_-)$ is the only resonant term and is much larger than other contributions, such that:
\begin{equation}
    S_\mathrm{II}^{(1)}[\omega] \approx n_\mathrm{het}\kappa_\mathrm{ex,L}n_\mathrm{th} |M(\omega_-)|^2.
\end{equation}
At this point, it is convenient to set the heterodyne frequency equal to the input frequency, or $\omega_-=0$, such that $\omega_-=\omega$. $M(\omega)$ can be simplified to
\begin{equation}
    M(\omega) \approx \frac{1}{\sqrt{2}}\frac{g \bar{a}_\mathrm{o}}{\omega + \Delta + J} \chi(\omega_-),
\end{equation}
given a combined answer of
\begin{equation}
     S_\mathrm{II,t}^{(1)}[\omega] \approx \frac{n_\mathrm{het}\kappa_\mathrm{ex,L}^2 n_\mathrm{th} n_\mathrm{in} g^2}{4} \left|\frac{\chi(\omega)}{(\Delta-J_\mathrm{O})(\omega+\Delta+J)}\right|^2.
\end{equation}
This transduced signal peaks at $\omega=\Omega$, $\mathrm{Re}(\Delta_\mathrm{L(R)})=\Omega$ and $J_\mathrm{O}=\Omega/2$, where it reads
\begin{equation}
    \max(S_\mathrm{II,t}^{(1)}[\omega]) = \frac{16 n_\mathrm{het}\kappa_\mathrm{ex,L}^2 n_\mathrm{th} n_\mathrm{in} g^2}{\Gamma \kappa^4}.
\end{equation}
Similarly, now assuming detuned modes $\Delta_\mathrm{R} \gg \Delta_\mathrm{L},J_\mathrm{O},\Omega$, we can simplify $M(\omega)$ to
\begin{equation}
    M(\omega) \approx \frac{-g \bar{a}_\mathrm{L}}{\omega + \Delta_\mathrm{L}} \chi(\omega),
\end{equation}
giving:
\begin{equation}
     S_\mathrm{II,d}^{(1)}[\omega] \approx \kappa_\mathrm{ex,L}^2 n_\mathrm{th} n_\mathrm{in} g^2 \left|\frac{\chi(\omega)}{\Delta_\mathrm{L}(\omega+\Delta_\mathrm{L})}\right|^2,
\end{equation}
with
\begin{equation}
    \max(S_\mathrm{II,d}^{(1)}) = \frac{16 n_\mathrm{het}\kappa_\mathrm{ex,L}^2 n_\mathrm{th} n_\mathrm{in} g^2}{\Gamma \kappa^2\Omega^2}. 
\end{equation}
Together, this gives an enhancement of linear transduction sideband power of:
\begin{equation}
    \mathcal{E}_\mathrm{lin} = \frac{\max(S_\mathrm{II,t}^{(1)})}{\max(S_\mathrm{II,d}^{(1)})} = \frac{\Omega^2}{\kappa^2}.
\end{equation}

Next, we analyse nonlinear transduction, starting with the tuned system. More precisely, we analyse fluctuations at $2\Omega_j$, assuming $\Omega_j$ the eigenfrequency of an odd mechanical mode with $g_{\mathrm{R},j}=-g_{\mathrm{L},j}=g$. Again, we assume $\Delta$ and $J_\mathrm{O}$ close to optimal, which is $\mathrm{Re}(\Delta)\approx J_\mathrm{O}+\Omega$, $J_\mathrm{O}\approx \Omega/2$, and we assume that we excite the even optical mode. In such case, the expression of $S_\mathrm{II}^{(2)}$ is dominated by two terms:
\begin{multline}
    S_\mathrm{II,t}^{(2)} \approx \frac{n_\mathrm{het}\kappa_\mathrm{ex,L}n_\mathrm{th}^2}{2\pi} \int d\omega^\prime  N_{j,j}(\omega_-,\omega^\prime) \big( N^\ast_{j,j}(\omega_-,\omega^\prime)\\ + N^\ast_{j,j}(\omega_-,\omega_- -\omega^\prime) \big) \\ \approx \frac{n_\mathrm{het}\kappa_\mathrm{ex,L}n_\mathrm{th}^2}{2\pi} \int d\omega^\prime  2|N_{j,j}(\omega_-,\omega^\prime)|^2.
\end{multline}
$N_{j,j}$ can be simplified with the above assumptions and by assuming only the even optical mode is driven, which in turn entails that $M_{\mathrm{L},j}(\omega) = -M_{\mathrm{R},j}(\omega)$, which means that linear transduction happens only on the odd optical mode. Together, we have
\begin{multline}
    S_\mathrm{II,t}^{(2)} [\omega] \approx \frac{n_\mathrm{het}\kappa_\mathrm{ex,L} n_\mathrm{th}^2 n_\mathrm{cav} g^4}{2\pi} \\ \int d\omega^\prime \left| \frac{\chi_j(\omega^\prime) \chi_j(\omega-\omega^\prime)}{(\omega + \Delta + J_\mathrm{O})(\omega -\omega^\prime +\Delta -J_\mathrm{O})}  \right|^2,
\end{multline}
where now we consider a constant amount of photons in the cavity $n_\mathrm{cav}$. 
By assuming that the denominator is roughly constant over the frequency range where $\chi(\omega)$ peaks, i.e. the optical properties change much more slowly with frequency than the mechanical properties, we can extract that
\begin{equation}
    \max(S_\mathrm{II,t}^{(2)}) \approx \frac{64 n_\mathrm{het}\kappa_\mathrm{ex,L} n_\mathrm{th}^2 n_\mathrm{cav} g^4}{\Gamma \kappa^4}.
\end{equation}
For the detuned case, we have
\begin{multline}
    S_\mathrm{II,d}^{(2)} [\omega] \approx \frac{n_\mathrm{het}\kappa_\mathrm{ex,L} n_\mathrm{th}^2 n_\mathrm{cav} g^4}{\pi} \\ \int d\omega^\prime \left| \frac{\chi_j(\omega^\prime) \chi_j(\omega-\omega^\prime)}{(\omega + \Delta_\mathrm{L})(\omega -\omega^\prime +\Delta_\mathrm{L})}  \right|^2,
\end{multline}
which gives
\begin{equation}
    \max(S_\mathrm{II,d}^{(2)}) \approx \frac{32 n_\mathrm{het}\kappa_\mathrm{ex,L} n_\mathrm{th}^2 n_\mathrm{cav} g^4}{\Gamma \kappa^2 \Omega^2}.
\end{equation}
Together, this yields a nonlinear enhancement of
\begin{equation}
    \mathcal{E}_\mathrm{qua}= \frac{\max(S_\mathrm{II,t}^{(2)})}{\max(S_\mathrm{II,d}^{(2)})} = 2\frac{\Omega^2}{\kappa^2}.
\end{equation}
Note that for the equations in the main text, we have set $n_\mathrm{het}=1$ to further simplify expressions.

\end{document}